\documentclass[aps,pra,amsmath,amssymb,twocolumn,showpacs]{revtex4}

\usepackage{amsthm}
\usepackage{bbm}
\usepackage{color}
\usepackage{graphicx}
\usepackage{subfigure}

\usepackage[english]{babel}

\newtheorem{definition}{Definition}
\newtheorem{proposition}{Proposition}

\begin{document}

\title{Quantum teleportation with identical particles}

\author{Ugo Marzolino}
\affiliation{Univerza v Ljubljani, Jadranska 19, SI-1000 Ljubljana, Slovenija \\
Physikalisches Institut, Albert-Ludwigs-Universit\"at Freiburg, Hermann-Herder-Stra\ss e 3, 79104 Freiburg, Deutschland}

\author{Andreas Buchleitner}
\affiliation{Physikalisches Institut, Albert-Ludwigs-Universit\"at Freiburg, Hermann-Herder-Stra\ss e 3, 79104 Freiburg, Deutschland}

\date{\today}

\begin{abstract}
We study teleportation with identical massive particles. Indistinguishability imposes that the relevant degrees of freedom to be teleported are not particles, but rather addressable orthogonal modes. We discuss the performances of teleportation under the constraint of conservation of the total number of particles. The latter inevitably decreases the teleportation fidelity. Moreover, even though a phase reference, given by the coupling to a reservoir, circumverts the constraint, it does not restore perfect deterministic teleportation. The latter is only achievable with some special resource entangled states and when the number of particles tends to infinity. Interestingly, some of such states are the many-particle atomic coherent states and the ground state of cold atoms loaded into a double well potential, which are routinely prepared in experiments.
\end{abstract}

\maketitle

\section{Introduction}

The field of quantum information theory has grown considerably in the last two decades, achieving important results on the characterization of quantum correlations and on their applicability to quantum information processing \cite{NielsenChuang,Mintert2005,Horodecci,Buhrman2010,Tichy2011}. Quantum protocols consist in two or more agents sharing quantum correlated states, which can help them to perform computational and technological tasks. The vast majority of the work in quantum information theory has been developed in the setting of distinguishable particles where each agent owns a given number of particles. However, quantum mechanics predicts that particles behave differently when they are identical \cite{Landau,Messiah}. The key point is that identical particles cannot be individually addressed: an agent cannot distinguish which particle he is manipulating.

Experimental realizations of quantum information processes consist of identical particles, e.g. photons, atoms in optical lattices, or electrons in solid state systems. For instance, ultracold atomic gases \cite{Leibfried2003,Morsch2006,Lewenstein2007,Bloch2008,Giorgini2008,Negretti2011} can be controlled with very high precision, and therefore are a promising arena for the study of many-body physics and applications in quantum information. Identical particles can be distinguished if appropriate degrees of freedom, e.g. spatial confinements, unambiguously characterize each particle. The properties in terms of these degrees of freedom, e.g. their individual positions, are employed to label each particle \cite{Herbut2001,Herbut2006,Tichy2013}, without being further manipulated. Based on this remark, many proposals suggest to encode states of distinguishable particles into systems of identical particles \cite{Ionicioiu2002,Riebe2004,Barrett2004,Negretti2011,Rohling2012,Underwood2012,Inaba2014}, in order to implement quantum information processing.

Nevertheless, if one aims to build an integrated architecture, it is difficult to effectively distinguish identical particles. In particular, it can be required to exploit all the accessible degrees of freedom, in order to encode the desired protocol, with no way to label or distinguish particles. For these reasons, the peculiarities of identical particles should be properly considered for a good definition of quantum correlations and for practical purposes of quantum information processing. In the present paper, we focus on the analysis of a specific quantum protocol, the generalization of teleportation \cite{Bennett1993} with massive, identical particles, and on the use of entangled states to improve achievable teleportation performances.

Entanglement, which is considered a key resource for teleportation, was debated for identical particles under several assumptions \cite{Zanardi2002-1,Schliemann2001,Paskauskas2001,Li2001,Ghirardi2002,Eckert2002,Wiseman2003,Ghirardi2004,Benatti2014-2}. We apply the approach developed in \cite{Zanardi2001,Barnum2004,Zanardi2004,Narnhofer2004,Benatti2010,Benatti2012,Benatti2014}, where entanglement is defined via non-classical correlations between subsets of observables. This is a very general and powerful approach that recovers the standard definition of entanglement for distinguishable particles, while for systems of identical particles it accounts for quantum correlations between occupations of orthogonal modes in the Fock space. Applying this framework to quantum information processing, each agent owns and locally manipulates modes, such as wells in optical lattices or hyperfine levels of molecular systems, which can be experimentally addressed \cite{Esteve2008,Wurtz2009,Gross2010,Riedel2010}. A key feature of massive particles, like atoms and constituents of condensed matter systems, is the conservation of the number of particles, mathematically described by a superselection rule \cite{Bartlett2007}. This feature is responsible for a very different behaviour of entanglement among identical particles as compared to the case of photons and distinguishable particles, manifest, for instance, in a simpler detectability of entanglement \cite{Benatti2012,Benatti2012-2,Benatti2014}, a high robustness against noise \cite{Benatti2012-2,Marzolino2013,Benatti2014}, and a different geometry of entangled states \cite{Benatti2012-2,Benatti2014}.

The properties of entanglement in its presently used definition have been studied in fermionic superconducting systems \cite{Zanardi2002-2,Vedral2004}, electrons in low-dimentional semiconductors \cite{Buscemi2006,Buscemi2007}, and bosonic ultracold gases \cite{Giorda2004,Anders2006,Goold2009,Benatti2010,Benatti2011,Argentieri2011,Cramer2013,Benatti2014}, and exploited in several applications, such as quantum data hiding \cite{Verstraete2003}, teleportation \cite{Marinatto2001,Schuch2004,Heaney2009,Molotkov2010}, Bell's inequalities \cite{Ashhab2009,Heaney2010}, dense coding \cite{Heaney2009}, and quantum metrology \cite{Benatti2010,Benatti2011,Argentieri2011,Benatti2014}. We shall discuss teleportation in detail, which plays a fundamental role in quantum computation \cite{Gottesman1999,Gross2007,Chiu2013}. The present analysis unifies and largely generalizes all the previous proposals of teleportation of identical particles, in the light of properties studied in \cite{Benatti2012,Benatti2012-2}.

In the original teleportation protocol \cite{Bennett1993} one agent, Alice, wants to teleport an arbitrary, perhaps unknown, state to another agent, Bob. Alice owns the state to be teleported and a share of a resource state, and Bob owns the remaining part of the resource state. The algorithm of the standard teleportation is the following: \emph{i)} Alice performs a projective measurement onto the basis of maximally entangled states of her states; \emph{ii)} Alice sends Bob the result of the measurement; \emph{iii)} Bob performs a suitable operation on his state, conditioned on the message he got from Alice. In the setting of distinguishable particles, if the shared state is a pure, maximally entangled state, Bob ends up with a state identical to the initial state to be teleported, while the state on Alice's side has been transformed by the measurement. Otherwise, Alice and Bob can distil maximally entangled states from a larger number of not maximally entangled states, in order to perform an optimal teleportation \cite{Bennett1996}. The teleportation can also be applied to a part of an entangled state. In this case, Alice initially owns a subsystem entangled with another subsystem in possession of a third party, and uncorrelated with the resource state and Bob's subsystem. After the teleportation, Alice's subsystem is entangled neither with Bob's nor with the third party subsystem, while the subsystems respectively owned by Bob and the third party become entangled. This application is called entanglement swapping, and can be useful for sharing entanglement at long distances, required for quantum networks \cite{Briegel1998,Kimble2008}.

This paper is organized as follows: In section \ref{entanglement}, we give the basic definition of entanglement and review some of its properties. In section \ref{twomode}, we introduce the teleportation protocol, where one mode of a two-mode state is teleported using a two-mode resource state. The conservation of the total number of particles for the independent operations performed during the teleportation protocol reduces the performance. This results in the impossibility of a perfect teleportation for any finite number of particles in the resource state. In section \ref{perf}, we analyse the teleportation efficiency with some exemplary resource states, and show that the efficiency grows with the number of particles in the resource state. These resource states include the maximally entangled states, but also states that can be more easily prepared, such as SU(2) coherent states \cite{Arecchi1972,Perelomov}, and the ground state of a double well potential with intrawell interactions \cite{Buonsante2012}. Additional details are shown in appendix \ref{perf.app}. We comment on the more cumbersome case of many-mode states to be teleported, as compared to two-mode states, in section \ref{multimode}. In section \ref{reference}, we prove the impossibility of perfect teleportation with a finite number of particles, for any teleportation protocol. As a byproduct, we prove that it is not possible to recover perfect teleportation even considering the usual way to overcome the particle number superselection rule, i.e. to couple the system to a reservoir \cite{Kitaev2004}. Finally, we summarize our conclusions in section \ref{discussions}. We use a mathematical style ({\bf Definition}, {\bf Proposition}), in order to emphasize a few key statements.

\section{Entanglement} \label{entanglement}

We now introduce basic definitions and notation which we will rely on in the rest of this paper. Specialist readers already fluent in the algebraic characterization of the entanglement of indistinguishable particles may skip this section and directly proceed to Section \ref{twomode}. We start with an algebraic framework which generalizes and recovers both entanglement of distinguishable particles and entanglement of identical particles \cite{Benatti2010,Benatti2012,Benatti2012-2,Benatti2014}. Let us consider a many-body system described by the Hilbert space ${\cal H}$. The algebra of all bounded operators, including all the observables, is denoted by ${\cal B}({\cal H})$\footnote{The algebra ${\cal{B}({\cal H})}$ is generated by differentiation of the so-called Weyl operators. Any function of the creation and annihilation operators is obtained by a proper differentiation of the Weyl operators \cite{EspositoMarmoSudarshan}.}. Unlike the usual definition \cite{Horodecci}, we move the attention from the states and the tensor product induced partitioning of ${\cal H}$ to the observables and local structures of ${\cal B}({\cal H})$.

\begin{definition}[Algebraic bipartition]
An \emph{algebraic bipartition} of ${\cal B}({\cal H})$ is any pair
$({\cal A}_1, {\cal A}_2)$ of commuting subalgebras,
${\cal A}_1, {\cal A}_2\subset {\cal B}({\cal H})$.
\end{definition}

\noindent
Any element of ${\cal A}_1$ commutes with any element of ${\cal A}_2$, $[{\cal A}_1, {\cal A}_2]=0$. The notion of locality lies in the commutativity of the subalgebras, ensuring the compatibility of observables of ${\cal A}_1$ with observables of ${\cal A}_2$.

\begin{definition}[Local operators]
An operator is said to be \emph{local} with respect to the bipartition $({\cal A}_1, {\cal A}_2)$, if it is the product $A_1 A_2$ of an operator 
$A_1$ in ${\cal A}_1$ and some $A_2$ in ${\cal A}_2$.
\end{definition}

\noindent
With the above definitions, we can now state the definition of separable and entangled states.

\begin{definition}[Entangled states] \label{ent.state}
A state $\rho$ is said to be \emph{separable} with respect to the bipartition $({\cal A}_1, {\cal A}_2)$ if the expectation of any local operator $A_1 A_2$ can be decomposed into a convex combination of
products of local expectations:

\begin{eqnarray}
& & {\rm tr}(\rho A_1 A_2)=\sum_k\lambda_k \, {\rm tr}\big(\rho_k^{(1)}A_1\big){\rm tr}\big(\rho_k^{(2)}A_2\big),\nonumber \\
& & 
\lambda_k\geq 0,\qquad \sum_k\lambda_k=1,
\end{eqnarray}

\noindent
with $\rho_k^{(1)}$ and $\rho_k^{(2)}$ admissible states of the system. Otherwise, the state is \emph{entangled}.
\end{definition}

Let us now focus on many-body systems whose constituents are $N$ bosons which fill $M$ different modes. The formalism of second quantization is more convenient for such systems. Let us introduce creation and annihilation operators $a^\dagger_j, a_j$, $j=1, 2,\ldots,M$, for each mode, with the bosonic commutation relations, $[a_j,\,a^\dagger_l]=\delta_{jl}$. The total Hilbert space ${\cal H}_N$ of the system is spanned by the Fock states

\begin{equation} \label{fock.states}
|k_1\rangle\otimes|k_2\rangle\otimes\cdots\otimes|k_M\rangle= \frac{(a_1^\dagger)^{k_1}\, (a_2^\dagger)^{k_2}\, \cdots\, (a_M^\dagger)^{k_M}\,|0\rangle}{\sqrt{k_1!\, k_2!\cdots k_M!}},
\end{equation}

\noindent
where the integer $k_j$ is the occupation number of the $j$-th mode such that $\sum_{j=1}^M k_j=N$. We use tensor product notation to write Fock states: for instance, $a_1^\dag a_2^\dag|0\rangle=a_1^\dag|0_1\rangle\otimes a_2^\dag|0_2\rangle$. We have to keep in mind that this tensor product structure is constrained by the conservation of the number of particles. The norm-closure of the set of polynomials in all creation and annihilation operators,
$\{a^\dagger_j,\, a_j\}$, $j=1,2,\dots, M$, is the algebra
${\cal B}({\cal H}_N)$. We define bipartitions of this algebra by splitting the set of creation and annihilation operators into two disjoint sets
$\{a_j^\dagger,a_j \, | \, j=1,2\dots,m\}$ and 
$\{a_j^\dagger,a_j \, | \, j=m+1,m+2,\dots,M\}$ with arbitrary $m$. The norm-closure of all polynomials in the creation and annihilation operators of the first (second) set is the subalgebra ${\cal A}_1$ (${\cal A}_2$). According to the previous Definition \ref{ent.state}, a pure state is $({\cal A}_1,{\cal A}_2)$-separable if and only if

\begin{equation}
|\psi\rangle={\cal P}(a^\dagger_1, \dots, a^\dagger_m)\cdot 
{\cal Q}(a^\dagger_{m+1},\ldots ,a^\dagger_M)\ |0\rangle,
\label{sep}
\end{equation}

\noindent
where ${\cal P}$ and ${\cal Q}$ are arbitrary functions. Mixed $({\cal A}_1,{\cal A}_2)$-separable states are convex combinations of pure $({\cal A}_1,{\cal A}_2)$-separable states. See \cite{Benatti2010,Benatti2012,Benatti2014} for a detailed analysis.

Since the total number of particles is conserved, all the observables and the density matrices commute with the total number operator. In other words, there is a superselection rule \cite{Bartlett2007} which forbids any coherent superpositions of states with different total numbers of particles. This is a key feature of such systems, and at the very heart of some recent results \cite{Benatti2012,Benatti2012-2}. Any orthonormal basis of $({\cal A}_1,{\cal A}_2)$-separable states can be relabelled as

\begin{eqnarray}
&& |k,\sigma\rangle\otimes|N-k,\sigma'\rangle, \qquad \sigma=1,2, \dots , D_k^{(m)}, \nonumber \\
&& \sigma'=1, 2, \dots , D_{N-k}^{(M-m)}, \qquad D_k^{(m)}={k+m-1 \choose k}. \qquad \label{sep.basis}
\end{eqnarray}

\noindent
The integer $k$ in Eq. \eqref{sep.basis} counts the number of particles in the first $m$ modes,
while $\sigma$ labels the different ways in which $k$ particles can fill those modes. Similarly, $\sigma'$ labels the ways in which the remaining $N-k$ particles can occupy the other $M-m$ modes. Any pure $({\cal A}_1,{\cal A}_2)$-entangled state is a coherent superposition of at least two $({\cal A}_1,{\cal A}_2)$-separable states.

There are two qualitatively different ways to superimpose separable states. The first is the superposition of states labeled by different $\sigma,\sigma'$, keeping $k$ fixed. Physically, the local numbers of particles are fixed in such states, thus the mode-bipartition corresponds to a particle-bipartition. A state which exhibits only these coherences has a block-diagonal structure in the label $k$:

\begin{equation} \label{block.diag}
\sum_{k=0}^N\sum_{\sigma,\tau=1}^{D_k^{(m)}}\sum_{\sigma',\tau'=1}^{D_{N-k}^{(M-m)}}\rho_{k \sigma\sigma', k\tau\tau'}\ |k,\sigma\rangle\langle k,\tau|\otimes|N-k,\sigma'\rangle\langle N-k,\tau'|.
\end{equation}

\noindent
This class of block-diagonal states includes all the states which are separable and positive under partial transposition (PPT) \cite{Benatti2012}. The set of states in each diagonal block, i.e. for each fixed value of $k$, is defined on the span of $\{|k,\sigma\rangle\otimes|N-k,\sigma'\rangle\}_{\sigma,\sigma'}$. This is a Hilbert subspace of dimension $D_k^{(m)}D_{N-k}^{(M-m)}$, isomorphic to the unconstrained tensor product space $\mathbbm{C}^{D_k^{(m)}}\otimes\mathbbm{C}^{D_{N-k}^{(M-m)}}$. Therefore, the mathematical features, and thus their physical consequences, are qualitatively analogous to those of distinguishable particles \cite{Benatti2012,Benatti2012-2}. The sum of the single block dimensions \cite{Prudnikov} is the dimension of the total Hilbert space

\begin{equation} \label{dim.gen}
\sum_{k=0}^ND_k^{(m)}D_{N-k}^{(M-m)}={N+M-1 \choose N}\equiv D.
\end{equation}
Because of the above isomorphism, such states are natural candidates for realizing quantum protocols developed for distinguishable particles with identical bosons \cite{Ionicioiu2002,Negretti2011,Rohling2012,Underwood2012,Inaba2014,
Byrnes2014}.

The second kind of superposition is the superposition of states labeled by different $k$, which is not compatible with a particle-bipartition. These states are not block-diagonal and, thus, not PPT \cite{Benatti2012}. The entanglement of such states is more powerful and robust than entanglement of distinguishable particles, because separable states in our framework are condensed to a zero measure subset. In fact, this kind of entanglement cannot be washed out by any mixtures with separable states, and mixtures with entangled states can erase such entanglement only under very special conditions \cite{Benatti2012-2}. On the other hand, local noisy dynamics cannot completely dissipate entanglement at finite times \cite{Marzolino2013}, contrary to what happens to distinguishable particles \cite{Zyczkowski2001,Carvalho2004,Fine2005}. All these phenomena reflect the fact that the set of separable states has zero measure.

Besides its definition, entanglement of identical particles can also be quantified with so-called entanglement measures. The theory of entanglement measures is based on the same principles of that for distinguishable particles \cite{Vidal2000,Horodecci}, provided that any physical state commutes with the total number of particles in our framework. Some entanglement measures for identical particles have been discussed in \cite{Schuch2004,Benatti2012,Benatti2012-2,Benatti2014}: negativity, robustness of entanglement and entanglement entropy. In accordance with these entanglement measures, we call a pure bipartite state of $N$ particles maximally entangled if its reduced density matrices are maximally mixed within the subspace with no more than $N$ particles. In the rest of this paper, we use negativity as entanglement measure.

In the following, we employ entangled states of identical particles, as defined above, as a resource for quantum teleportation. We start with discussing a teleportation protocol that uses two-mode states. The single block dimensions of two-mode states are $D_k^{(m)}D_{N-k}^{(M-m)}=1$. Thus, the only contribution to entanglement is due to coherent superpositions among different local particle numbers $k$.

\section{Teleportation protocol with two-mode states} \label{twomode}

We disuss the teleportation of one mode of a two-mode state $|\psi_{12}\rangle$ with the help of a two-mode, shared resource state $\rho_{34}$. The labels $1,2,3,4$ number the modes. The initial global state is $|\psi_{12}\rangle\langle\psi_{12}|\otimes\rho_{34}$, where

\begin{equation} \label{initial.state}
|\psi_{12}\rangle=\sum_{k=0}^N c_k|k\rangle_1\otimes|N-k\rangle_2, \qquad \sum_{k=0}^N|c_k|^2=1,
\end{equation}

\noindent
and $\rho_{34}$ is a general two-mode state with $\nu$ particles

\begin{equation} \label{res}
\rho_{34}=\sum_{k,l=0}^\nu\left(\rho_{34}\right)_{k,l}|k\rangle_3{\,}_3\langle l|\otimes|\nu-k\rangle_4{\,}_4\langle \nu-l|.
\end{equation}

\noindent
The second and the third modes are owned by the sender, Alice. She aims to teleport the state of the second mode to the target mode, the forth one, owned by the receiver, Bob. It does not matter whether the first mode is in possession of Alice or not.

First, if $\nu\geqslant N$ Alice performs a projective measurement of her modes, that involves projectors $P_{23}^{(l,\lambda)}=|\phi_{23}^{(l,\lambda)}\rangle\langle\phi_{23}^{(l,\lambda)}|$, onto the following orthogonal states

\begin{eqnarray} \label{states.meas}
&& |\phi_{23}^{(l,\lambda)}\rangle=\sum_{k=0}^N\frac{e^{2\pi i\frac{\lambda k}{N+1}}}{\sqrt{N+1}} \, |N-k\rangle_2\otimes|k+l\rangle_3, \nonumber \\
&& l\in\{0,1,\dots,\nu-N\}, \qquad \lambda\in\{0,1,\dots,N\},
\end{eqnarray}

\noindent
(where the phases $2\pi i\frac{\lambda k}{N+1}$ ensure the orthogonality).

We assess the entanglement and completeness of the basis \eqref{states.meas} for a comparison with teleportation protocols for distinguishable particles, because there a complete projective measurement onto maximally entangled states is required for perfect teleportation \cite{Bennett1993}. The states \eqref{states.meas} have the same amount of entanglement as the maximally entangled states of $N$ two-mode particles. Moreover, they span a $(\nu-N+1)(N+1)$-dimensional Hilbert space. Nevertheless, since the second (third) mode can be filled at most by $N$ ($\nu$) particles, all the possible pure states of these modes span an $(\nu+1)(N+1)$-dimensional Hilbert space. In other words, the conservation of the total number of particles applies to each state that is prepared independently, i.e. $|\psi_{12}\rangle$ and $\rho_{34}$, and not to the reduced state of the second and the third mode which have interacted with the other modes. On the other hand, Alice performs the measurement independently from the first and the fourth mode, thus the projections on the states (\ref{states.meas}) must commute with the total number of particles. Therefore, the states (\ref{states.meas}) do not define a complete measurement on the second and the third mode. Additional projectors onto the missing subspace should be considered, in order to make the measurement without post-selection trace preserving. However, the additional projectors cannot project onto states with the same amount of entanglement.

We will consider the complete set of orthogonal projectors $P_{23}^{(l,\lambda)}=|\phi_{23}^{(l,\lambda)}\rangle\langle\phi_{23}^{(l,\lambda)}|$ onto the following basis

\begin{eqnarray} \label{states.meas.2}
&& |\phi_{23}^{(l,\lambda)}\rangle=\sum_{k=\max\{0,-l\}}^{\min\{N,\nu-l\}}\frac{e^{2\pi i\frac{\lambda k}{{\cal C}_l}}}{\sqrt{\cal C}_l}|N-k\rangle_2\otimes|k+l\rangle_3, \nonumber \\
&& l\in\{-N,-N+1\dots,\nu\}, \quad \lambda\in\{0,1,\dots,{\cal C}_l-1\}, \quad
\end{eqnarray}

\noindent
where we have defined the function

\begin{equation}
{\cal C}_l=
\begin{cases}
N+l+1 & \mbox{if } -N\leqslant l\leqslant 0 \\
N+1 & \mbox{if } 0\leqslant l\leqslant \nu-N \\
\nu-l+1 & \mbox{if } \nu-N\leqslant l\leqslant \nu
\end{cases}
\end{equation}
which equals the cardinality of the sum in equation \eqref{states.meas.2}. The projection onto the states \eqref{states.meas.2} take into account also the cases when $0\leqslant\nu<N$. The records of Alice's measurement are $(l,\lambda)$ which label the outcome corresponding to the projection onto each basis state \eqref{states.meas.2}. If Alice records $(l,\lambda)$, the state changes into

\begin{widetext}
\begin{eqnarray} \label{op.telep}
& & \mathbbm{1}_1\otimes P_{23}^{(l,\lambda)}\otimes\mathbbm{1}_4\big(|\psi_{12}\rangle\langle \psi_{12}|\otimes\rho_{34}\big)\mathbbm{1}_1\otimes P_{23}^{(l,\lambda)}\otimes\mathbbm{1}_4= \nonumber \\
& & =\sum_{k,j=\max\{0,-l\}}^{\min\{N,\nu-l\}}\left(\rho_{34}\right)_{k+l,j+l}c_k\bar c_j \, \frac{e^{2\pi i\lambda\frac{j-k}{{\cal C}_l}}}{{\cal C}_l} \, |k\rangle_1{\,}_1\langle j|\otimes|\phi_{23}^{(l,\lambda)}\rangle\langle\phi_{23}^{(l,\lambda)}|\otimes|\nu-k-l\rangle_4{\,}_4\langle\nu-j-l|,
\end{eqnarray}
\end{widetext}

\noindent
where $\mathbbm{1}_j=\sum_{k\geqslant 0}|k\rangle_j{\,}_j\langle k|$ is the identity operator on the $j$-th mode, and $\{c_k\}_k$ are the coefficients of the initial state \eqref{initial.state}.

At this point the state of Bob's mode depends on the outcome of Alice's measurement. In order to reconstruct the state to be teleported, Alice sends Bob the outcome $(l,\lambda)$ of her measurement, via a classical channel, and Bob subsequently applies the operation $V_4^{(l,\lambda)}\rho (V_4^{(l,\lambda)})^\dag$ to the fourth mode, where

\begin{equation} \label{V_4}
V_4^{(l,\lambda)}=\sum_{k=\max\{0,-l\}}^{\min\{N,\nu-l\}} e^{2\pi i\frac{\lambda k}{{\cal C}_l}}|N-k\rangle_4{\,}_4\langle \nu-k-l|.
\end{equation}

Since the conservation of the total number of particles is central in our analysis, we now show that the operations $V_4^{(l,\lambda)}$ are fully consistent with this conservation law. Indeed, $V_4^{(l,\lambda)}$ can be implemented by a unitary operation which preserves the total particle number on the forth mode and on an additional fifth mode:

\begin{eqnarray}
&& V_4^{(l,\lambda)}\rho (V_4^{(l,\lambda)})^\dag={\rm tr}_5(\tilde V_{45}^{(l,\lambda)}\rho_4\otimes|\kappa_l\rangle_5{\,}_5\langle\kappa_l|(\tilde V_{45}^{(l,\lambda)})^\dag), \nonumber \\
&& |\kappa_l\rangle_5=\frac{(a_5^\dag)^{\kappa_l}}{\sqrt{\kappa_l!}}|0\rangle_5, \qquad \kappa_l\geqslant\min\{0,\nu-N-l\},
\end{eqnarray}

\noindent
where ${\rm tr}_5$ is the trace over the fifth mode, and

\begin{widetext}
\begin{eqnarray}
\tilde V_{45}^{(l,\lambda)} & = & \sum_{k=\max\{0,-l\}}^{\min\{N,\nu-l\}} U_k^{(l,\lambda)}+\mathbbm{1}_{(l,\lambda)}, \nonumber \\
U_k^{(l,\lambda)} & = &
\begin{cases}
\displaystyle e^{2\pi i\frac{\lambda k}{{\cal C}_l}}|N-k\rangle_4{\,}_4\langle N-k|\otimes|\kappa_l\rangle_{5}{\,}_5\langle\kappa_l| & \textnormal{if} \quad l=\nu-N \\
\displaystyle e^{2\pi i\frac{\lambda k}{{\cal C}_l}}|N-k\rangle_4{\,}_4\langle\nu-k-l|\otimes|\kappa_l+\nu-N-l\rangle_{5}{\,}_5\langle \kappa_l|+{\rm h.c.} & \textnormal{if} \quad l\neq\nu-N
\end{cases}.
\end{eqnarray}
\noindent
$\mathbbm{1}_{(l,\lambda)}$ is the identity matrix on the subspace orthogonal to the support of each $U_k^{(l,\lambda)}$. Thus, $\tilde V_{45}^{(l,\lambda)}$ is a unitary transformation which commutes with the total number of particles.

Now, we come to the analysis of the final state and how similar to the original state \eqref{initial.state} to be teleported it is. Since Bob knows the outcome $(l,\lambda)$ on Alice's side, and accordingly transforms his state, he ends up with the mode $4$ of one of the states $\rho_{14}^{(l,\lambda)}$ with probability $p_{(l,\lambda)}$, as given by the partial trace over the second and the third mode

\begin{eqnarray} \label{telep.term}
p_{(l,\lambda)}\rho_{14}^{(l,\lambda)} & \equiv & {\rm tr}_{23}\left[\mathbbm{1}_1\otimes P_{23}^{(l,\lambda)}\otimes V_{4}^{(l,\lambda)}\big(|\psi_{12}\rangle\langle\psi_{12}|\otimes\rho_{34}\big)\mathbbm{1}_1\otimes P_{23}^{(l,\lambda)}\otimes(V_{4}^{(l,\lambda)})^\dag\right]= \nonumber \\
& = & \sum_{k,j=\max\{0,-l\}}^{\min\{N,\nu-l\}}\left(\rho_{34}\right)_{k+l,j+l}\frac{c_k\bar c_j}{{\cal C}_l}|k\rangle_1{\,}_1\langle j|\otimes|N-k\rangle_4{\,}_4\langle N-j|,
\end{eqnarray}
with ${\rm tr}\big(\rho_{14}^{(l,\lambda)}\big)=1$. The average of the teleported state over all possible outcomes on Alice's side, which we will use to discuss the performances of the above protocol, is given by the following local operations with classical communication (LOCC):

\begin{equation} \label{teleported}
\mathcal{T}\big[|\psi_{12}\rangle\langle\psi_{12}|\big]=\sum_{l=-N}^\nu\sum_{\lambda=0}^{{\cal C}_l-1}p_{(l,\lambda)}\rho_{14}^{(l,\lambda)}=\sum_{l=-N}^{\nu}\sum_{k,j=\max\{0,-l\}}^{\min\{N,\nu-l\}} c_k\bar c_j\left(\rho_{34}\right)_{k+l,j+l}|k\rangle_1{\,}_1\langle j|\otimes|N-k\rangle_4{\,}_4\langle N-j|.
\end{equation}
\end{widetext}
We stress that the state (\ref{teleported}) is not Bob's final state, but the average over all possible final states at Bob's end.

Following \cite{Horodecki1999}, the faithfulness of the teleportation is quantified by the fidelity

\begin{equation} \label{fid}
f=\int d\psi \langle\psi|\mathcal{T}\big[|\psi\rangle\langle\psi|\big]|\psi\rangle,
\end{equation}
where $d\psi$ is the uniform distribution over all pure states, and the states $|\psi\rangle$ and the average teleported state $\mathcal{T}\big[|\psi\rangle\langle\psi|\big]$ stem from the same Hilbert space. The teleportation fidelity is the average overlap between the final state of the first and fourth mode and the initial state of the first and second mode, thus measures how similar these states are. In order to define the uniform distribution, consider an arbitrary state $|\psi\rangle=\sum_{k=0}^N c_k|e_k\rangle$ with $\sum_k|c_k|^2=1$, in a Hilbert space spanned by the orthonormal basis $\{|e_k\rangle\}_{k=0,\dots,N}$. Re-writing the coefficients $c_k=r_k e^{i\varphi_k}$, with $r_k\geq 0$ and $0\leq\varphi_k<2\pi$, the uniform distribution over the pure states is induced by the Haar measure of the group of the unitary transformations \footnote{Other distributions of pure states $d\psi$ can be considered. For instance, a flat distribution in the variables $r_k$ and $\phi_k$ has been studied in \cite{Facchi2006}. The only differences are some numerical factors, while the scaling with $N$ does not change. We focus on the measure (\ref{measure}) for comparison with the results in \cite{Horodecki1999}.}:

\begin{equation} \label{measure}
d\psi=\frac{N!}{\pi^{N+1}}\delta\left(1-\sum_{k=0}^N r_k^2\right)\prod_{k=0}^N r_k dr_k d\varphi_k,
\end{equation}

\noindent
see e.g. \cite{Facchi2008,Facchi2010-2}. For instance, the average values of $|c_k|^\alpha$ and $|c_k|^\alpha |c_j|^\beta$ with $k\neq j$ are

\begin{eqnarray}
\label{av1} \int d\psi \, |c_k|^\alpha & = & \frac{\Gamma(1+\frac{\alpha}{2})\Gamma(N+1)}{\Gamma(N+1+\frac{\alpha}{2})},
\\
\label{av2} \int d\psi \, |c_k|^\alpha |c_j|^\beta & = & \frac{\Gamma(1+\frac{\alpha}{2})\Gamma(1+\frac{\beta}{2})\Gamma(N+1)}{\Gamma(N+1+\frac{\alpha+\beta}{2})},
\end{eqnarray}
where $\Gamma$ is the Euler's gamma function \cite{AbramowitzStegun} and $\alpha,\beta>-2$. Inserting equation \eqref{teleported} for the average teleported state into the definition \eqref{fid} of the fidelity, and using the averages \eqref{av1} and \eqref{av2}, we derive the following expression

\begin{widetext}
\begin{equation} \label{fidelity}
f=\sum_{l=-N}^\nu\sum_{k,j=\max\{0,-l\}}^{\min\{N,\nu-l\}}\left(\rho_{34}\right)_{k+l,j+l}\int d\psi |c_k|^2|c_j|^2=\frac{2}{N+2}\left(1+\sum_{k\neq j; \, k,j=0}^\nu\frac{\max\big\{0,N+1-|k-j|\big\}}{2(N+1)}\left(\rho_{34}\right)_{k,j}\right),
\end{equation}
\end{widetext}
where the function $\max$ comes from counting the number of times each term $\left(\rho_{34}\right)_{k,j}$ appears within the triple sum inherited by equation \eqref{teleported} \footnote{Note that each term $\left(\rho_{34}\right)_{k',j'}$, with $k'=k+l$ and $j'=j+l$, appears in the triple sum of equation \eqref{teleported} $N+1$ times if $k=j$, $N$ times if $|k-j|=1$, $N-1$ times if $|k-j|=2$, ... only once if $|k-j|=N$, and never if $|k-j|>N$. In order to see this, look at the sum over $l$ as a squared $(N+1)\times(N+1)$ shadow shifting along the diagonal blocks of $\rho_{34}$. The matrix elements of $\rho_{34}$ covered at each sweep of the shadow are $\left(\rho_{34}\right)_{k+l,j+l}$ for $k$ and $j$ contributing to the sum of \eqref{teleported}.}.

If the agents are not interested in teleporting the state itself, but in sharing as much entanglement as possible between the first and the fourth mode, then a more relevant figure of merit is the measure of entanglement of the final state between the first and the fourth mode. In general, the final state is mixed, and the negativity \cite{Zyczkowski1998,Vidal2002} is the most easily computable measure of entanglement for mixed states:

\begin{equation}
{\cal N}(\rho)=\frac{{\rm tr}\sqrt{(\rho^T)^2}-1}{2},
\end{equation}
where $^T$ denotes partial transposition \cite{Peres1996}. While in general it vanishes for some entangled states, the negativity of a two-mode state with a fixed number of particles is a faithful measure of entanglement \cite{Argentieri2011,Benatti2012}. Given a two-mode state $\rho$, its negativity is

\begin{equation}
{\cal N}(\rho)=\frac{1}{2}\sum_{k\neq j}\big|\rho_{k,j}\big|.
\end{equation}
We consider the double average of the negativity over the outcomes $(l,\lambda)$ and the uniform distribution of the initial state as a quantifier of the final entanglement:

\begin{eqnarray} \label{av.ent}
E & = & \int d\psi\sum_{l=-N}^\nu\sum_{\lambda=0}^{{\cal C}_l-1}p_{(l,\lambda)}{\cal N}(\rho_{14}^{(l,\lambda)}) \nonumber \\
& = & \frac{\pi}{8}\sum_{k\neq j; \, k,j=0}^\nu\frac{\max\big\{0,N+1-|k-j|\big\}}{N+1}\big|\left(\rho_{34}\right)_{k,j}\big|, \qquad
\end{eqnarray}

\noindent
where the function $\max$ has the same origin as in equation \eqref{fidelity}. Equation \eqref{av.ent} ranges between zero and $\pi N/8$. The upper bound $E\leq\pi N/8$ is proven by noting that the teleportation protocol does not act on the first mode and cannot increase the entanglement between the first mode and the rest. Thus, the average final entanglement $E$ is not larger than the average entanglement over all pure initial states, namely $\int d\psi \, {\cal N}\big(|\psi\rangle\langle\psi|\big)=\pi N/8$. The entanglement of each final state $\rho_{14}^{(l,\lambda)}$ is independent of whether the local operation $V_4^{(l,\lambda)}$ has been performed. We notice that

\begin{equation} \label{triangle}
\frac{8E}{\pi}\geq(N+2)f-2
\end{equation}
follows from the triangle inequality for the absolute value applied to \eqref{av.ent}, and from the positivity of the fidelity $f$ \eqref{fidelity}. We will use this inequality to translate properties of the fidelity \eqref{fidelity} to properties of the average teleported entanglement \eqref{av.ent}.

Let us briefly comment on the case of a non-entangled state $|\psi_{12}\rangle$ instead of the state \eqref{initial.state}. This case is not of practical interest for two reasons. First, it does not allow for the protocol of entanglement swapping. Second, the perfect teleportation of the second mode of such a state is possible without any entangled resource state. Indeed, the non-entangled two-mode states are only the Fock states in the choosen basis of modes \cite{Benatti2010,Benatti2012}. In particular, the states of the second mode are the Fock states $|\bar k\rangle_2$, with $c_k=\delta_{k,N-\bar k}$. In order to broadcast the information of this state, it is enough that Alice measures the number operator $a_2^\dag a_2$, which provides the full information on $\bar k$, and communicates the result to Bob. Then, Bob can prepare locally the state $|\bar k\rangle_4$. It is however instructive to look at the fidelity of the teleportation protocol in this case. Equation (\ref{fidelity}) with $c_k=\delta_{k,N-\bar k}$ tells us that the state $|\bar k\rangle_2$ cannot be teleported if $l<\bar k-N$ or $l>\nu-N+\bar k$. Thus, the teleportation protocol \eqref{res}-\eqref{teleported.sep} fails to teleport any state of the Fock basis of the second mode, and the linearity of the protocol implies the following property:

\begin{proposition} \label{prop1}
The above teleportation protocol \eqref{res}-\eqref{teleported.sep} cannot perfectly teleport an arbitrarily entangled state $|\psi_{12}\rangle$. The teleportation fidelity \eqref{fidelity} is $f<1$.
\end{proposition}

One can wonder whether there is a better protocol which achieves $f=1$. The answer is negative, as we prove below, even considering the most general class of teleportation protocols. Before discussing generalizations of the above teleportation, we analyse the performances of the protocol presented in this section with some exemplary resource states.

\section{Teleportation performances} \label{perf}

In this section, we discuss the teleportation performances quantified by the fidelity and the average final entanglement of several interesting resource states, i.e. maximally entangled states, N00N states, SU(2) coherent states, and the ground state of a double well potential, as regards their relevance in other physical processes. Additional details are exposed in appendix \ref{perf.app}. We identify the useful states for teleportation, namely those states that outperform the separable states when employed for the teleportation protocol. The additional contribution to the fidelity, with respect to that of separable states, can be negative, unlike the contribution to the average final entanglement. Moreover, not all of the off-diagonal entries are relevant for teleportation. Indeed, the entries $(\rho_{34})_{k,j}$ with $|k-j|>N$ do not enter in the formula of the fidelity (\ref{fidelity}) and of the average final entanglement (\ref{av.ent}).

\subsection{Separable resources}

As mentioned above, separable resource states, namely

\begin{equation} \label{sep.res}
\rho_{34}=\sum_{k=0}^\nu \left(\rho_{34}\right)_{k,k} |k\rangle_3\langle k|\otimes|\nu-k\rangle_4\langle\nu-k|,
\end{equation}
provide threshold performances that define useful resource states. The teleportation fidelity \eqref{fidelity} and the average entanglement \eqref{av.ent} are the same for all resource states of this type \eqref{sep.res}:

\begin{eqnarray}
\label{fid.sep} f_{\rm sep} & = & \frac{2}{N+2}, \\
\label{ent.sep} E_{\rm sep} & = & 0.
\end{eqnarray}

\noindent
It is worth mentioning that $f_{\rm sep}$ equals the highest teleportation fidelity achievable with resource states of distinguishable $(N+1)$-level systems, which are separable or positive under partial transposition \cite{Horodecki1999}. This latter observation is consistent since resource states with a fixed number of particles are special cases of unconstrained states, e.g. photonic modes, which in turn are mathematically equivalent to states of distinguishable $(N+1)$-level systems.

\subsection{Maximally entangled resources and probabilistic, perfect teleportation} \label{max.ent.res}

Let us now consider the pure, maximally entangled resource state $\rho_{34}=|\phi_{34}\rangle\langle\phi_{34}|$ of $\nu$ two-mode particles \cite{Benatti2012,Benatti2012-2}, where

\begin{equation} \label{max.ent.state}
|\phi_{34}\rangle=\frac{1}{\sqrt{\nu+1}}\sum_{k=0}^\nu|k\rangle_3\otimes|\nu-k\rangle_4.
\end{equation}
If Alice's measurement results in a projection onto \eqref{states.meas.2}, with $0\leqslant l\leqslant\nu-N$, the state of the second mode is perfectly teleported to the fourth mode. These outcomes occur with probability $\frac{\nu-N+1}{\nu+1}$, see appendix \ref{perf.app}.

The direct computation of the teleportation fidelity (\ref{fidelity}) with elementary summations leads to

\begin{equation} \label{fid.max.ent}
f_{\rm max \, ent}=1-\frac{N}{3(\nu+1)}
\end{equation}

\noindent
which is always larger than the fidelity $f_{\rm sep}$ of separable resources, and is arbitrary close to one if $\nu\gg N$. Under this condition, the probability of the outcome $(l,\lambda)$ with $l<0$ or $l>\nu-N$ is negligible, and the initial state is almost perfectly teleported. The average entanglement \eqref{av.ent} of the teleported state is straightforwardly computed:

\begin{equation} \label{ent.max.ent}
E_{\rm max \, ent}=\frac{\pi N(3\nu-N+1)}{24(\nu+1)}
\end{equation}
which is smaller than the average entanglement $\pi N/8$ over all pure states, and converges to this value for $\nu\gg N$.

\subsection{N00N states}
\label{N00N.states}

Exemplary entangled resource states which are {\em not} useful for teleportation are the so-called N00N sates with $\nu$ particles:

\begin{equation} \label{N00N}
|\nu00\nu\rangle_{34}=\frac{1}{\sqrt{2}}\big(|\nu\rangle_3\otimes|0\rangle_4+|0\rangle_3\otimes|\nu\rangle_4\big), \qquad \nu>N.
\end{equation}

\noindent
From the direct computation of equations \eqref{fidelity} ad \eqref{av.ent}, they provide the same teleportation performances as separable resource states \eqref{sep.res}. See appendix \ref{perf.app} for the case $\nu\leqslant N$, which slightly outperforms separable resource states without qualitative improvement.

These features of N00N states are remarkable, because they are the most useful states in the high accuracy estimation of the relative phase between the two arms of an atomic interferometer \cite{Bollinger1996,Hyllus2010}. The apparent imbalance of their performance for different purposes can be explained because coherence among all Fock states, thus strong mode-entanglement, is needed to implement accurate teleportation, whereas only coherence between two Fock states with highly unbalanced population in two modes, thus weak mode-entanglement, is required for precise phase estimation. Indeed, accurate phase estimation is achieved with states that sensitively vary under phase shifts, and corresponds to a Heisenberg-like relation where the larger the variance of relative occupation number, the smaller the accuracy of the phase. Moreover, the counterpart of N00N states in first quantization, i.e. the so-called GHZ states, are used to perfectly teleport states of distinguishable particles in low dimensional Hilbert spaces \cite{Zhao2004,Marinatto2000,Banerjee2011,Saha2012}.

\subsection{SU(2) coherent states} \label{SU2.coh.states}

States studied in the context of mean field approximation and mesoscopic quantum coherent phenomena \cite{Raghavan1999,Benatti2009} are the so-called SU(2) coherent states \cite{Perelomov}, also known as atomic coherent states \cite{Arecchi1972}. These are states where all particles occupy the same combination of two modes defined by the population probabilities of both modes, $\xi$ and $1-\xi$, and the relative phase $\vartheta$:

\begin{widetext}
\begin{equation}
|\xi,\vartheta\rangle_{34}=\frac{1}{\sqrt{\nu!}}\left(\sqrt{\xi} \, e^{-i\frac{\vartheta}{2}}a_3^\dag+\sqrt{1-\xi} \, e^{i\frac{\vartheta}{2}}a_4^\dag\right)^\nu |0\rangle_3\otimes|0\rangle_4=\sum_{k=0}^\nu\sqrt{{\nu \choose k}}\xi^{\frac{k}{2}}\left(1-\xi\right)^{\frac{\nu-k}{2}} e^{i\vartheta\left(\frac{\nu}{2}-k\right)}|k\rangle_3\otimes|\nu-k\rangle_4.
\label{def.coh}
\end{equation}

\begin{figure}[htbp]
\centering
\subfigure[]{
\label{fid.sym.coh.a}
\includegraphics[width=0.48\columnwidth]{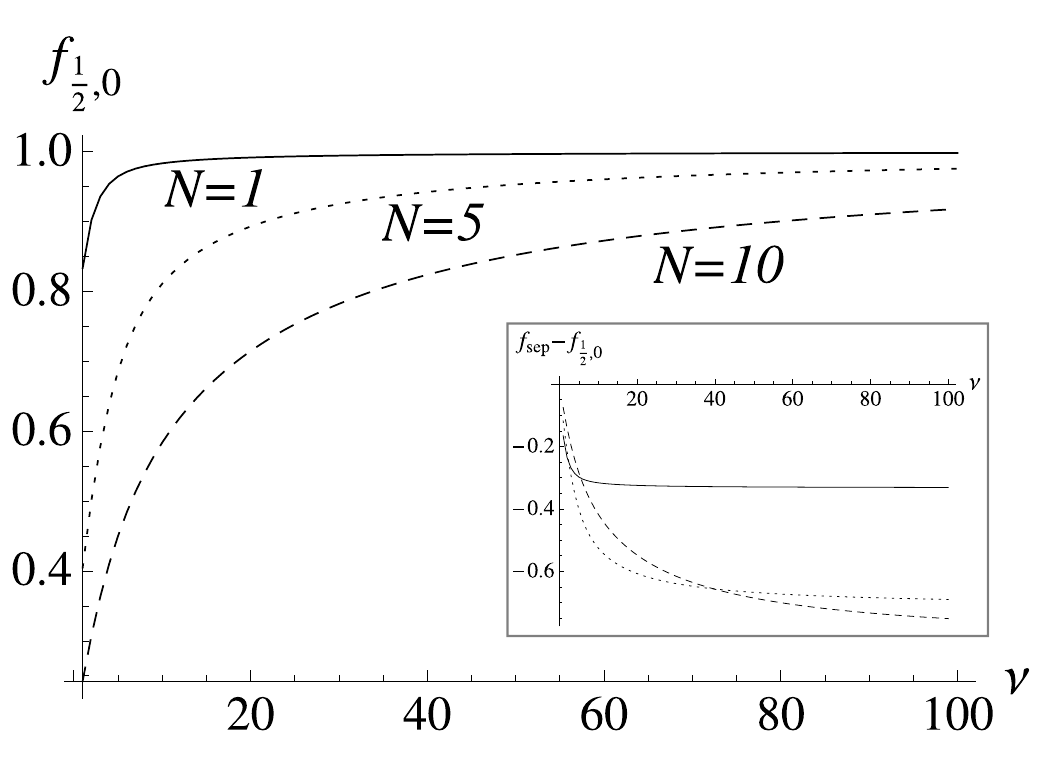}
}
\subfigure[]{
\label{fid.sym.coh.b}
\includegraphics[width=0.48\columnwidth]{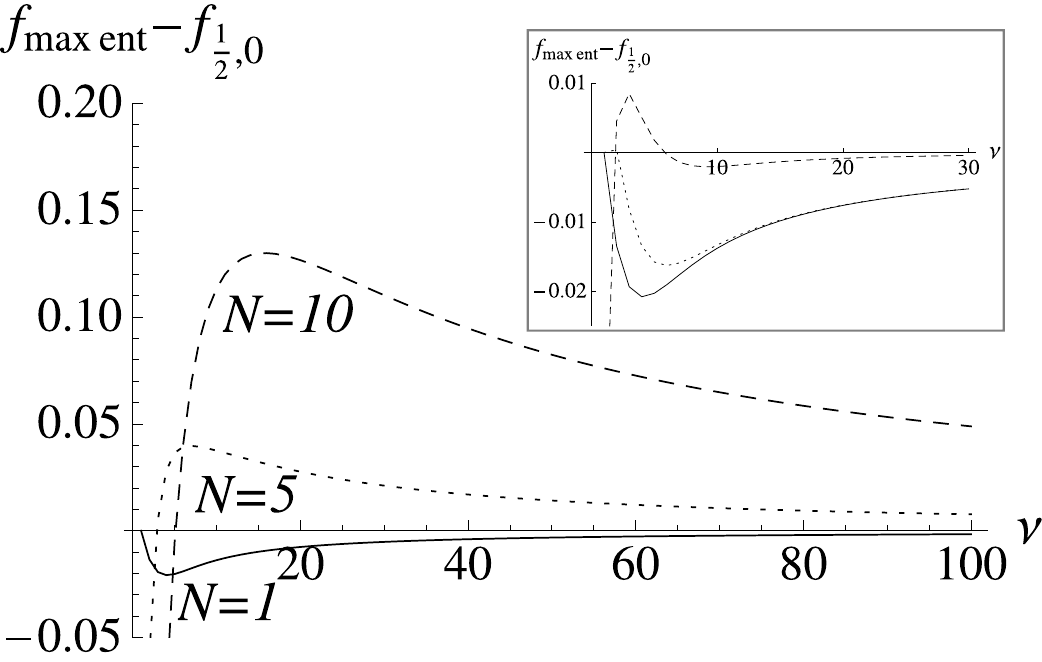}
}
\caption{Fidelity $f_{\frac{1}{2},0}$, equation \eqref{fidelity}, of the teleportation protocol with the resource state being the SU(2) symmetric coherent state $|\xi=1/2,\vartheta=0\rangle$, equation \eqref{def.coh}. Panel (a) shows the teleportation fidelity $f_{\frac{1}{2},0}$, the inset of (a) shows the difference $f_{\rm sep}-f_{\frac{1}{2},0}$ with respect to the fidelity with separable resource states \eqref{sep.res}, and panel (b) is the difference $f_{\rm max \, ent}-f_{\frac{1}{2},0}$ with respect to the fidelity with the maximally entangled resource state \eqref{max.ent.state}, for $N=1$ (continuous line), $N=5$ (dotted line), and $N=10$ (dashed line). The inset of (b) shows the cases where the fidelity with the SU(2) symmetric coherent resource state is larger than the fidelity with the maximally entangled resource state \eqref{max.ent.state} even for large $\nu$: $N=1$ (continuous line), $N=2$ (dotted line), $N=3$ (dashed line)}
\label{fid.sym.coh}
\end{figure}
\end{widetext}

We numerically compute the fidelity $f_{\frac{1}{2},0}$ and the average final entanglement $E_{\frac{1}{2},0}$ of the teleportation protocol when the resource state is a symmetric coherent state $|\xi=1/2,\vartheta=0\rangle_{34}$, i.e. with balanced population probability $\xi=1/2$, that is routinely prepared in the laboratory \cite{Gross2010,Riedel2010}. We plot teleportation performances and compare them with the performances of the maximally entangled resource state \eqref{max.ent.state}, with those of separable resource states \eqref{sep.res}, and with their maximum values in figures \ref{fid.sym.coh} and \ref{ent.sym.coh}. Teleportation performances are always better than performances of separable states, and can be very close to their maximum values for large $\nu$. In such limit, the binomial distribution in the definition \eqref{def.coh} of the coherent state becomes more and more flat, and approaches the uniform distribution that characterizes the superposition of the maximally entangled state.

\begin{widetext}

\begin{figure} [htbp]
\centering
\subfigure[]{
\label{ent.sym.coh.a}
\includegraphics[width=0.48\columnwidth]{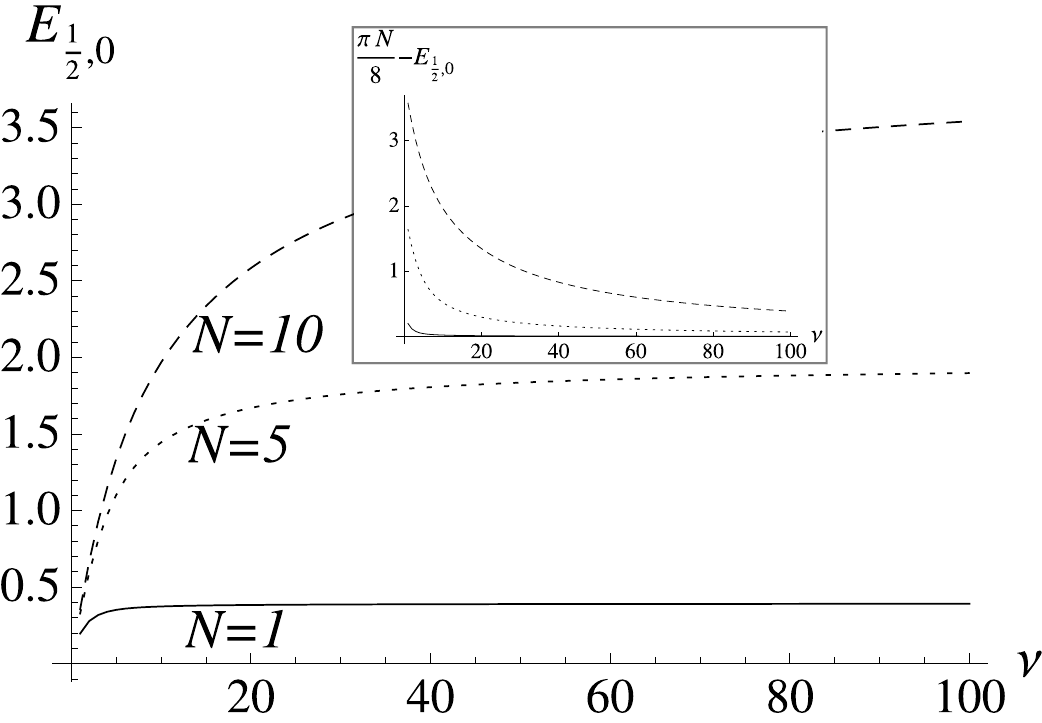}
}
\subfigure[]{
\label{ent.sym.coh.b}
\includegraphics[width=0.48\columnwidth]{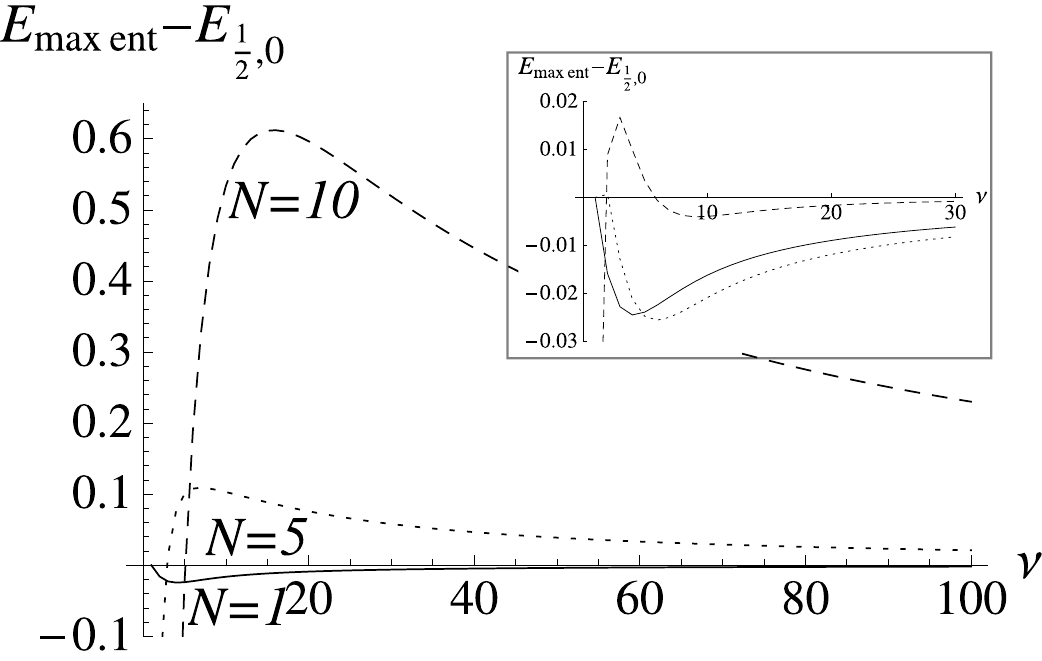}
}
\caption{Average final entanglement $E_{\frac{1}{2},0}$, equation \eqref{av.ent}, of the teleportation protocol with the resource state being the SU(2) symmetric coherent state $|\xi=1/2,\vartheta=0\rangle_{34}$, equation \eqref{def.coh}. Panel (a) shows the average final entanglement $E_{\frac{1}{2},0}$, the inset of (a) shows the difference $\frac{\pi N}{8}-E_{\frac{1}{2},0}$ with respect to the maximum average final entanglement, and panel (b) is the difference $E_{\rm max \, ent}-E_{\frac{1}{2},0}$ with respect to the average final entanglement with the maximally entangled resource state \eqref{max.ent.state}, for $N=1$ (continuous line), $N=5$ (dotted line), and $N=10$ (dashed line). The inset of (b) shows the cases where the average final entanglement with the SU(2) symmetric coherent resource state is larger than the average final entanglement with the maximally entangled resource state \eqref{max.ent.state} even for large $\nu$: $N=1$ (continuous line), $N=2$ (dotted line), $N=3$ (dashed line)}
\label{ent.sym.coh}
\end{figure}

\begin{figure} [htbp]
\centering
\centering
\subfigure[]{
\label{fid.coh}
\includegraphics[width=0.48\columnwidth]{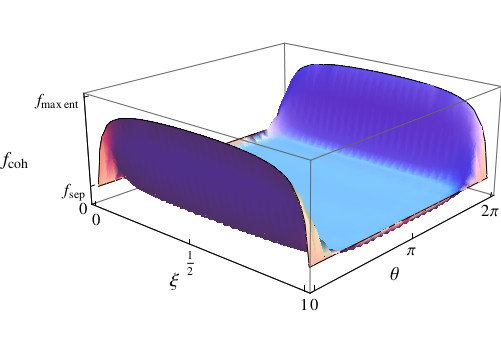}
}
\subfigure[]{
\label{ent.coh}
\includegraphics[width=0.48\columnwidth]{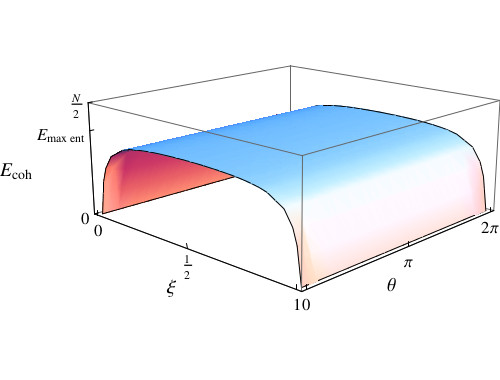}
}
\caption{Fidelity $f_{\rm coh}$, equation \eqref{fidelity} and panel (a), and average final entanglement $E_{\rm coh}$, equation \eqref{av.ent} and panel (b), of the teleportation protocol with the resource state being the SU(2) coherent state $|\xi,\vartheta\rangle$, equation \eqref{def.coh}, for $N=10$ and $\nu=100$. The fidelity $f_{\rm coh}$ is maximized by the symmetric coherent state $|\xi=1/2,\vartheta=0\rangle_{34}$, and quickly decays with $\vartheta$, while $E_{\rm coh}$ is maximixed by $\xi=1/2$ for any phase $\vartheta$. Nevertheless, phases $\vartheta\neq0$ can be compensated by local unitary operations (see appendix \ref{perf.app}).}
\label{perf.coh}
\end{figure}
\end{widetext}

We also plot the teleportation fidelity $f_{\rm coh}$ and the average final entanglement $E_{\rm coh}$ which can be generated with the help of a general with a general SU(2) coherent state, with fixed $N=10$ and $\nu=100$ in figure \ref{perf.coh}. Maximum fidelity is produced with the symmetric for the symmetric coherent state $|\xi=1/2,\vartheta=0\rangle$, and maximum average final entanglement is achieved for the symmetric coherent state $|\xi=1/2,\vartheta\rangle$ for any $\vartheta$. The average final entanglement does not depend on the phase $\vartheta$, since only the modulus of the off-diagonal elements of the density matrix enters in its definition \eqref{av.ent}.

Recall that perfect teleportation with identical particles, meaning fidelity $f=1$, is impossible, while for distinguishable particles it is attained with maximally entangled resource states. In this context, it is remarkable that some coherent states can attain almost perfect teleportation, and that they can on average even outperform maximally entangled resource states, in some parameter regimes. Indeed, coherent states are easy to prepare in experiments, see e.g. \cite{Gross2010,Riedel2010}, and \cite{Diehl2008} for a theoretical proposal of dissipative preparation.

In order to stress the application of the notion of entanglement discussed in this paper, notice that SU(2) coherent states are entangled and they are useful resources for teleportation. In this context, SU(2) coherent states never outperform classical metrology for phase estimation \cite{Wineland1994,Benatti2010,Benatti2011,Argentieri2011,Benatti2014}. Thus, teleportation with identical particles shows the effects and the usefulness of mode-entanglement, even for states where phase estimation does not.

\subsection{Ground states of the double well potential}

In this section, we discuss the ground state of the two-mode Bose-Hubbard Hamiltonian \cite{Lewenstein2007,Buonsante2012} as a resource for the teleportation protocol. This application is potentially relevant because this state can be prepared with present days technology, such as magnetic traps and evaporative cooling \cite{Thomas2002}. The Bose-Hubbard hamilonian

\begin{equation} \label{BH}
H=-\tau\left(a_3^\dag a_4+a_4^\dag a_3\right)+U\left((a_3^\dag)^2a_3^2+(a_4^\dag)^2a_4^2\right)
\end{equation}

\noindent
is fixed by the tunneling amplitude $\tau$ between the two well sites and the on-site interaction strength $U$.

Let us consider this ground state as the resource state of the teleportation protocol $\rho_{34}=|\textnormal{gs}_{34}\rangle\langle\textnormal{gs}_{34}|$, where

\begin{equation} \label{gs}
|\textnormal{gs}_{34}\rangle=\sum_{k=0}^\nu g_k|k\rangle_3\otimes|\nu-k\rangle_4.
\end{equation}

\noindent
There are four regimes parametrized by the ratio \cite{Buonsante2012}

\begin{equation} \label{gamma}
\gamma=\frac{\nu U}{\tau}:
\end{equation}

\begin{enumerate}
\item If $\gamma\gg-\sqrt{\nu}$ or $\gamma\ll\nu^2$, the tunneling term can be treated as a perturbation, and the ground state emerges as a superposition of few Fock states. With a similar analysis as compared to that in section \ref{N00N.states}, we can argue that the resulting state has poor entantanglement and is not very useful for teleportation.

\item If $-1+\nu^{-2/3}\ll\gamma\ll\nu^2$, then the ground state is a Gaussian superposition

\begin{equation} \label{gauss}
g_k=\frac{e^{-\frac{(k-\frac{\nu}{2})^2}{4\sigma_{\gamma}^2}}}{\sqrt{Z}}, \qquad \sigma_{\gamma}^2=\frac{\nu}{4\sqrt{\gamma+1}},
\end{equation}

\noindent
with the normalization

\begin{equation}
Z=\sum_{k=0}^\nu e^{-\frac{(k-\frac{\nu}{2})^2}{2\sigma_{\gamma}^2}}\simeq\frac{\nu}{2}\sqrt{2\pi\sigma_{\gamma}^2},
\end{equation}

\noindent
for large $\nu$. This range of $\gamma$ shall be called {\it single Gaussian regime}.

\item In a different regime, characterized by $-\sqrt{\nu}\ll\gamma\ll-1-\nu^{-2/3}$, the ground state is a superposition of two Gaussians \cite{Buonsante2012}:

\begin{eqnarray} \label{2gauss}
&& g_k=\frac{1}{\sqrt{Z'}}\left(e^{-\frac{\left(k-\frac{\nu}{2}-\frac{\nu}{2}\sqrt{1-\frac{1}{\gamma^2}}\right)^2}{4\sigma_{\gamma}'^2}}+e^{-\frac{\left(k-\frac{\nu}{2}+\frac{\nu}{2}\sqrt{1-\frac{1}{\gamma^2}}\right)^2}{4\sigma_{\gamma}'^2}}\right), \nonumber \\
&& \sigma_{\gamma}'^2=\frac{\nu}{4|\gamma|\sqrt{\gamma^2-1}},
\end{eqnarray}

\noindent
with the normalization

\begin{eqnarray}
Z' & = & \sum_{k=0}^\nu\left(e^{-\frac{\left(k-\frac{\nu}{2}-\frac{\nu}{2}\sqrt{1-\frac{1}{\gamma^2}}\right)^2}{4\sigma_{\gamma}'^2}}+e^{-\frac{\left(k-\frac{\nu}{2}+\frac{\nu}{2}\sqrt{1-\frac{1}{\gamma^2}}\right)^2}{4\sigma_{\gamma}'^2}}\right)^2 \nonumber \\
& \simeq & \nu\sqrt{2\pi\sigma_{\gamma}'^2}
\end{eqnarray}

\noindent
for large $\nu$. We shall refer to this instance as {\it double Gaussian regime}.

\item The transition range between the single and the double Gaussian regimes is the {\it critical regime} $-1-\nu^{-2/3}\ll\gamma\ll-1+\nu^{-2/3}$. In the critical regime, the ground state is a superposition peaked around $k=\nu/2$, but less confined than a Gaussian since it starts to split into two imbalanced occupations of the sites.
\end{enumerate}

\begin{widetext}

\begin{figure}[htbp]
\centering
\subfigure[]{
\label{fid.gauss.a}
\includegraphics[width=0.48\columnwidth]{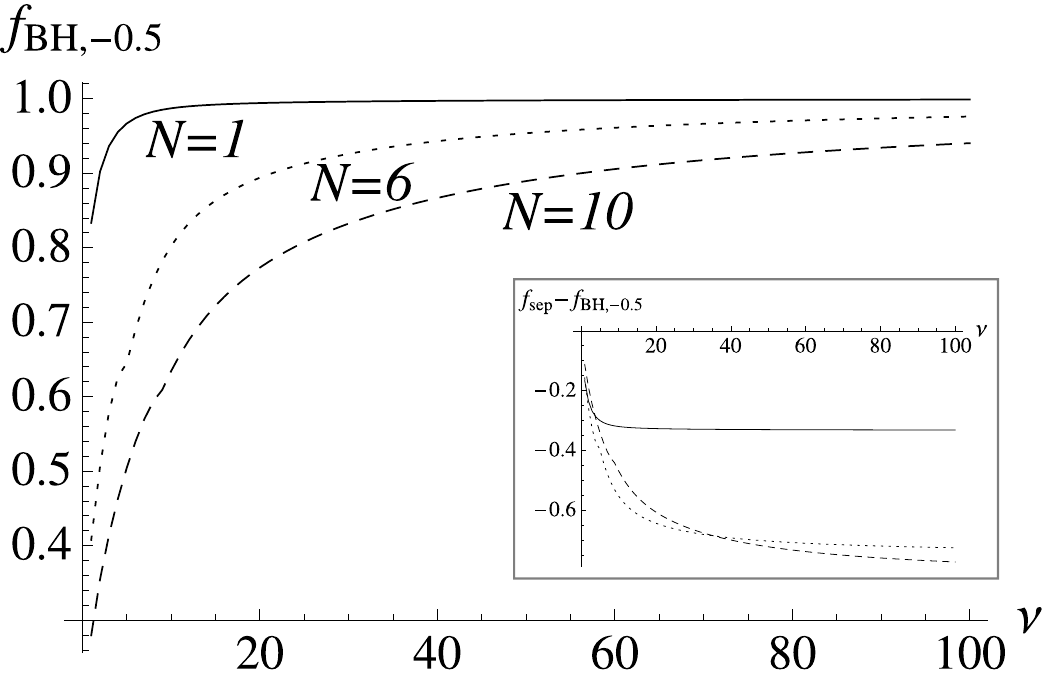}
}
\subfigure[]{
\label{fid.gauss.b}
\includegraphics[width=0.48\columnwidth]{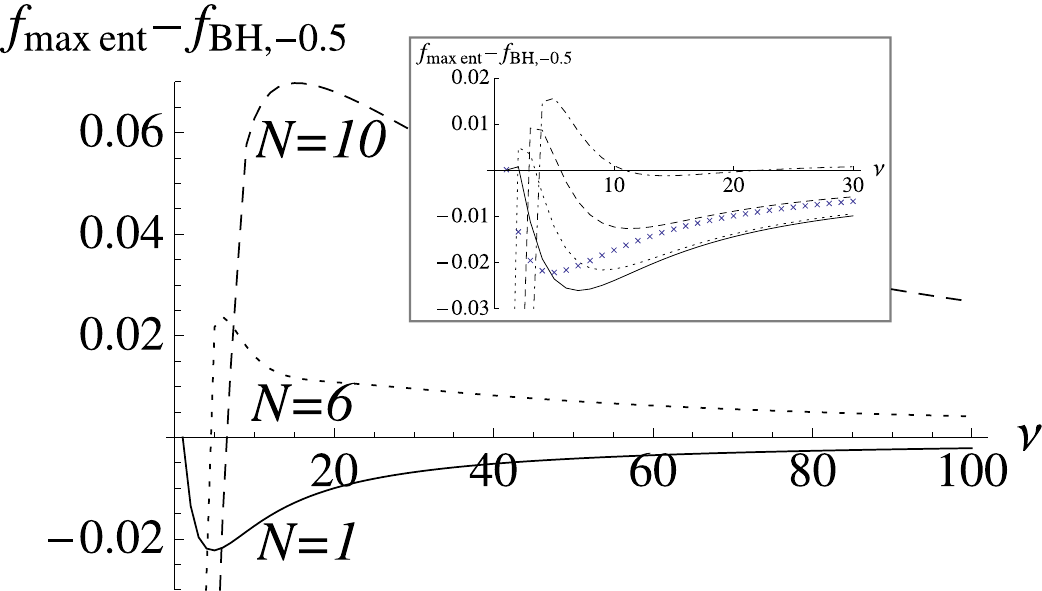}
}
\caption{Fidelity $f_{{\rm BH},-0.5}$, equation \eqref{fidelity}, of the teleportation protocol with the resource state being the ground state of the Hamiltonian (\ref{BH}) in the single Gaussian regime for $\gamma=-0.5$, equations \eqref{gs} and \eqref{gauss}. Panel (a) shows the teleportation fidelity $f_{{\rm BH},-0.5}$, the inset of (a) shows the difference $f_{\rm sep}-f_{{\rm BH},-0.5}$ with respect to the fidelity with separable resource states \eqref{sep.res}, and panel (b) is the difference $f_{\rm max \, ent}-f_{{\rm BH},-0.5}$ with respect to the fidelity with the maximally entangled resource state \eqref{max.ent.state}, for $N=1$ (continuous line), $N=6$ (dotted line), and $N=10$ (dashed line). The inset of (b) shows the cases where the fidelity with the Gaussian resource state \eqref{gauss} is larger than the fidelity with the maximally entangled resource state \eqref{max.ent.state} even for large $\nu$: $N=1$ (crosses), $N=2$ (continuous line), $N=3$ (dotted line), $N=4$ (dashed line), $N=5$ (dash-dotted line).}
\label{fid.gauss}
\end{figure}

\begin{figure} [htbp]
\centering
\subfigure[]{
\label{ent.gauss.a}
\includegraphics[width=0.48\columnwidth]{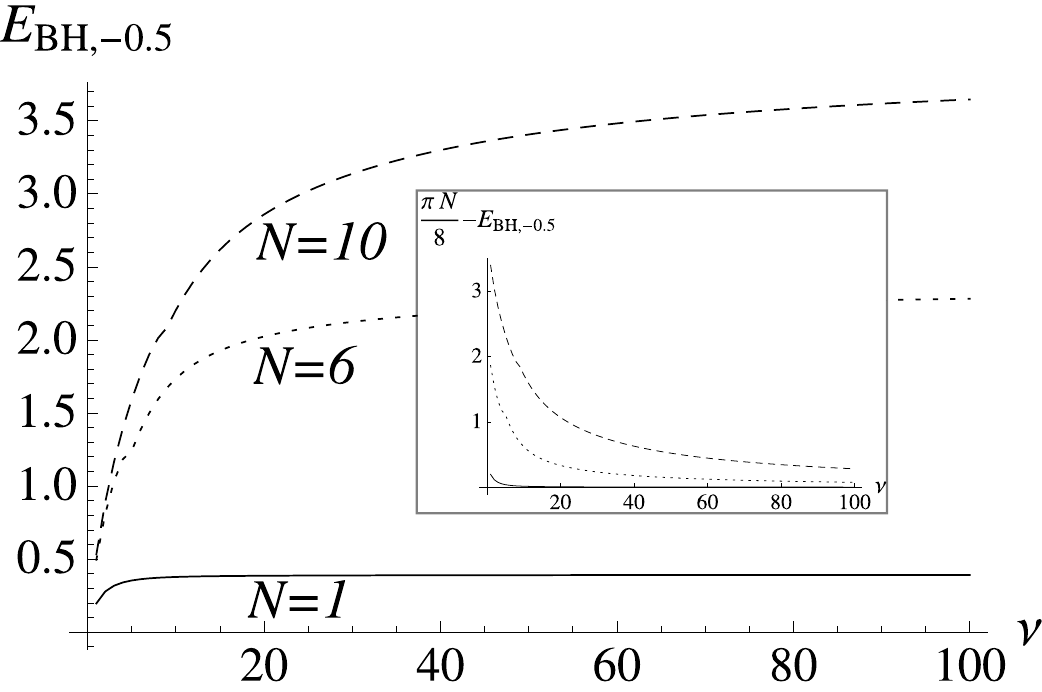}
}
\subfigure[]{
\label{ent.gauss.b}
\includegraphics[width=0.48\columnwidth]{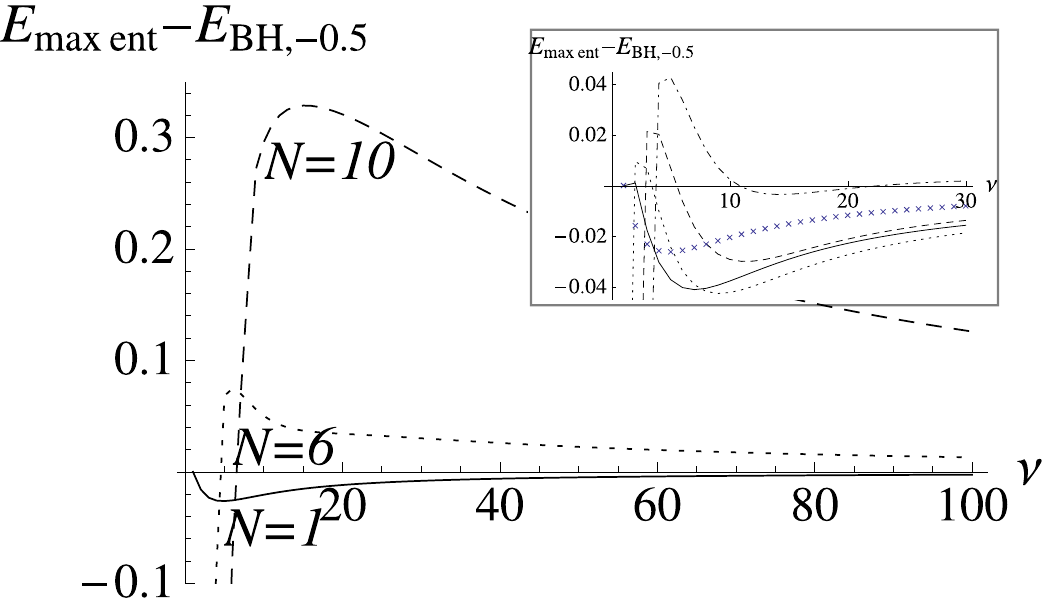}
}
\caption{Average final entanglement $E_{{\rm BH},-0.5}$, equation \eqref{av.ent}, of the teleportation protocol with the resource state being the ground state of the Hamiltonian (\ref{BH}) in the single Gaussian regime for $\gamma=-0.5$, equations \eqref{gs} and \eqref{gauss}. Panel (a) shows the average final entanglement $E_{{\rm BH},-0.5}$, the inset of (a) shows the difference $\frac{\pi N}{8}-E_{{\rm BH},-0.5}$  with respect to the maximum average final entanglement, and panel (b) is the difference $E_{\rm max \, ent}-E_{{\rm BH},-0.5}$ with respect to the average final entanglement with the maximally entangled resource state \eqref{max.ent.state}, for $N=1$ (continuous line), $N=5$ (dotted line), and $N=10$ (dashed line). The inset of (b) shows the cases where the average final entanglement with the Gaussian resource state \eqref{gauss} is larger than the average final entanglement with the maximally entangled resource state \eqref{max.ent.state} even for large $\nu$: $N=1$ (crosses), $N=2$ (continuous line), $N=3$ (dotted line), $N=4$ (dashed line), $N=5$ (dash-dotted line).}
\label{ent.gauss}
\end{figure}
\end{widetext}

Considering the ground state of the Bose-Hubbard model as a resource for the teleportation protocol, we numerically compute the fidelity $f_{{\rm BH},\gamma}$, equation \eqref{fidelity}, and the average final entanglement $E_{{\rm BH},\gamma}$, equation \eqref{av.ent}. We plot teleportation performances and compare them with the performances of the maximally entangled resource state \eqref{max.ent.state}, with those of separable resource states \eqref{sep.res}, and with their maximum values in figures \ref{fid.gauss} and \ref{ent.gauss}. We choose $\gamma=-0.5$ because we do not observe any qualitative difference within the single and the double Gaussian regimes. Teleportation performances are always better than those of separable resource states, and can be very close to their maximum values for large $\nu$ because, as for the symmetric coherent state, the wave function of the ground state spreads and approaches the uniform distribution realized by the superposition in the maximally entangled state \eqref{max.ent.state}. In figures \ref{fid.BH} and \ref{ent.BH}, we plot teleportation performances as functions of $\gamma$, and comparisons with those of the maximally entangled resource state \eqref{max.ent.state}.

We notice that there is a region in both regimes where the fidelity $f_{{\rm BH},\gamma}$ and the average final entanglement $E_{{\rm BH},\gamma}$ are larger than the corresponding quantities of the maximally entangled state. When $N$ grows, this region shrinks around the boundary with the critical regime $\gamma\sim-1\pm \nu^{-2/3}$. Thus, the single Gaussian and the double Gaussian approximations may be no longer reliable and these numerical results may not necessarily match the exact behaviour of the ground state of (\ref{BH}). Neverthless, (\ref{gauss}) and (\ref{2gauss}) are {\it bona fide} states which outperform the average performances of the maximally entangled states, even if they do not coincide with the ground state of the double well potential.

\begin{widetext}

\begin{figure} [htbp]
\centering
\subfigure[]{
\label{fid.BH.a}
\includegraphics[width=0.48\columnwidth]{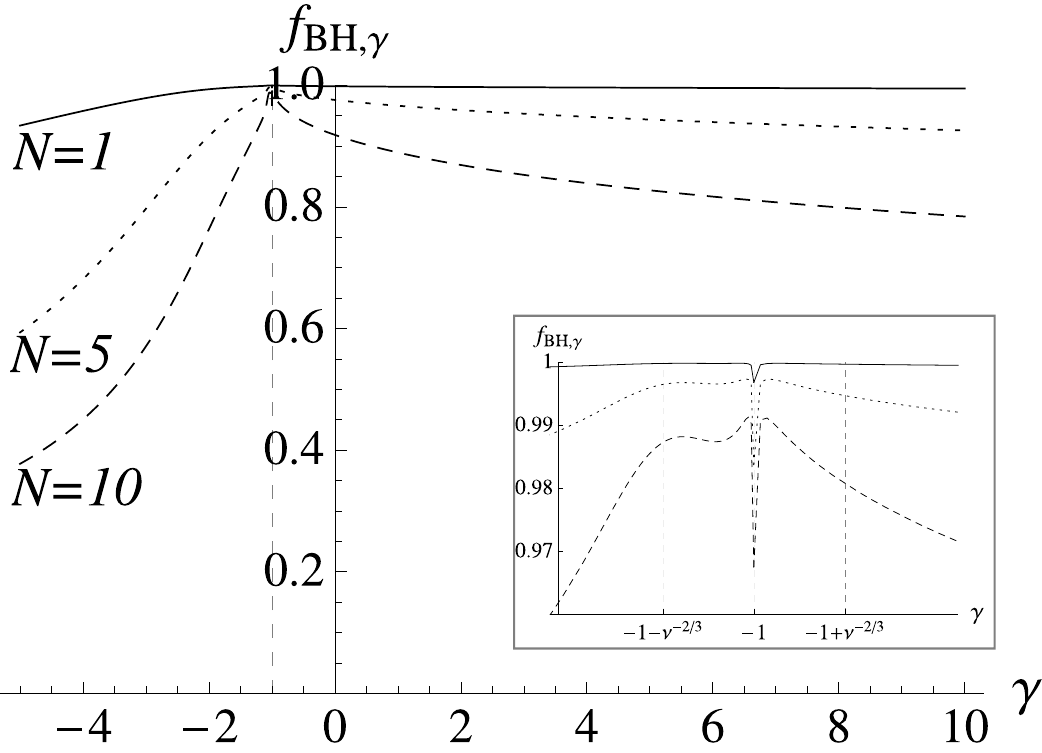}
}
\subfigure[]{
\label{fid.BH.b}
\includegraphics[width=0.48\columnwidth]{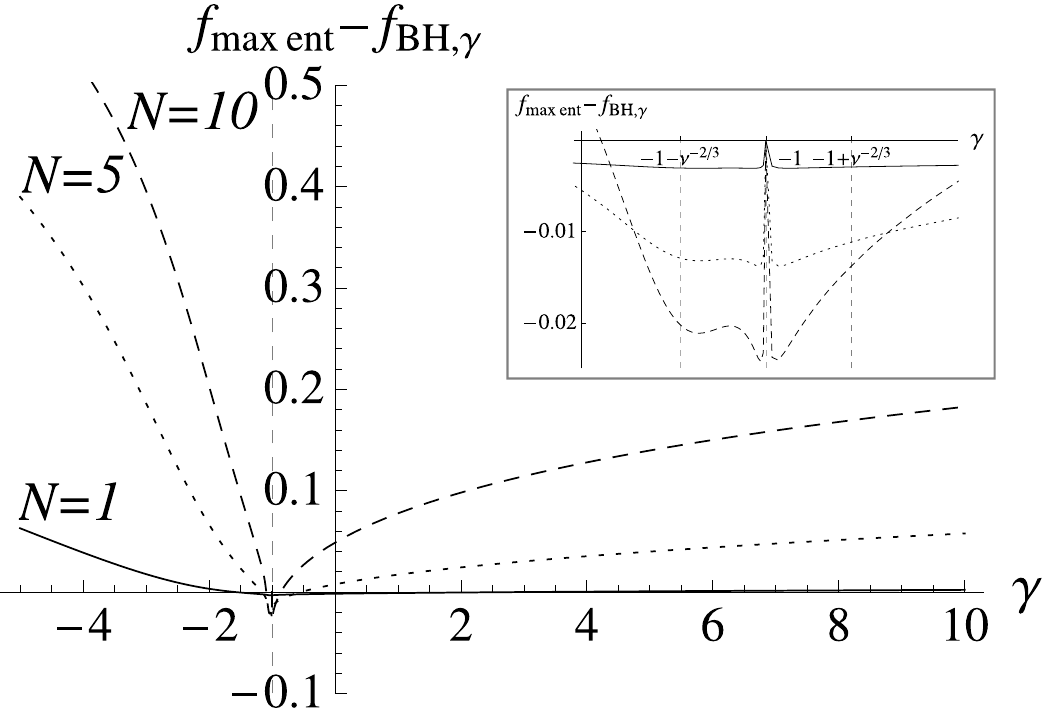}
}
\caption{Fidelity $f_{{\rm BH},-0.5}$, equation \eqref{fidelity}, of the teleportation protocol with the resource state being the ground state of the Hamiltonian (\ref{BH}) in the single Gaussian regime for $\nu=100$, equations \eqref{gs} and \eqref{gauss}. Panel (a) shows the teleportation fidelity $f_{{\rm BH},-0.5}$, and panel (b) is the difference $f_{\rm max \, ent}-f_{{\rm BH},-0.5}$ with respect to the fidelity with the maximally entangled resource state \eqref{max.ent.state}, for $N=1$ (continuous line), $N=5$ (dotted line), and $N=10$ (dashed line). The insets show the zoom of the critical region $\left|\gamma+1\right|\leqslant\nu^{-2/3}$. Since the ground state of \eqref{BH} is not known in the critical region, the resource state of the teleportation fidelity plotted in the above insets is not the ground state of \eqref{BH}. It is instead the continuation of the ground state from the outside region $\left|\gamma+1\right|\geqslant\nu^{-2/3}$.}
\label{fid.BH}
\end{figure}

\begin{figure} [htbp]
\centering
\subfigure[]{
\label{ent.BH.a}
\includegraphics[width=0.48\columnwidth]{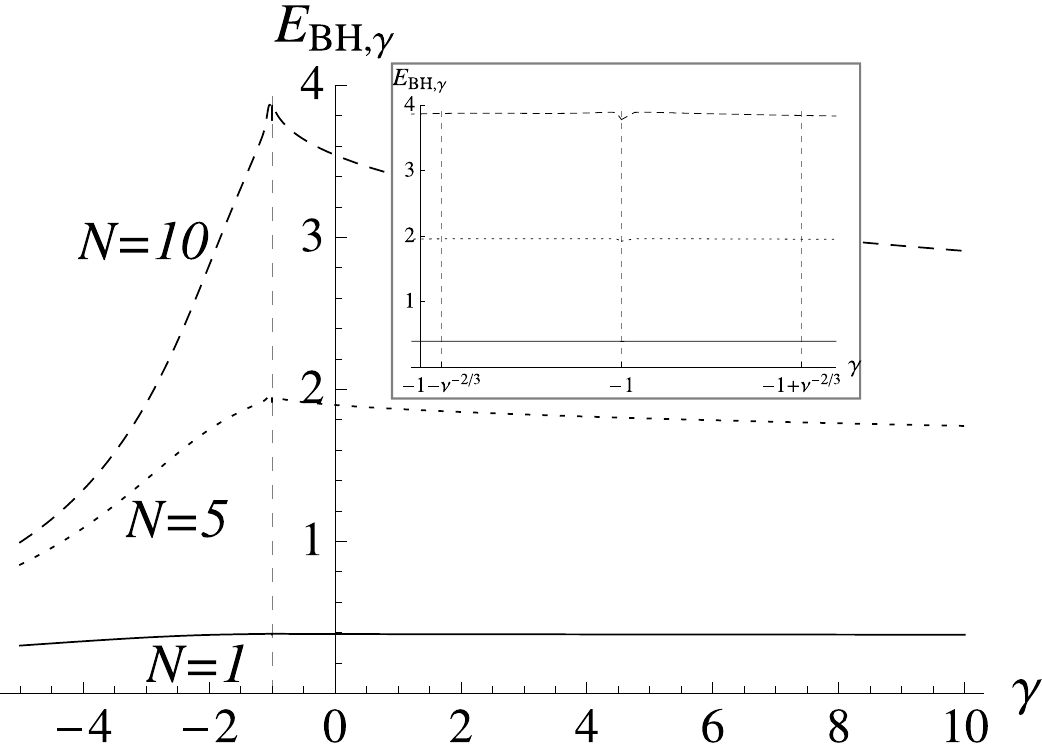}
}
\subfigure[]{
\label{ent.BH.b}
\includegraphics[width=0.48\columnwidth]{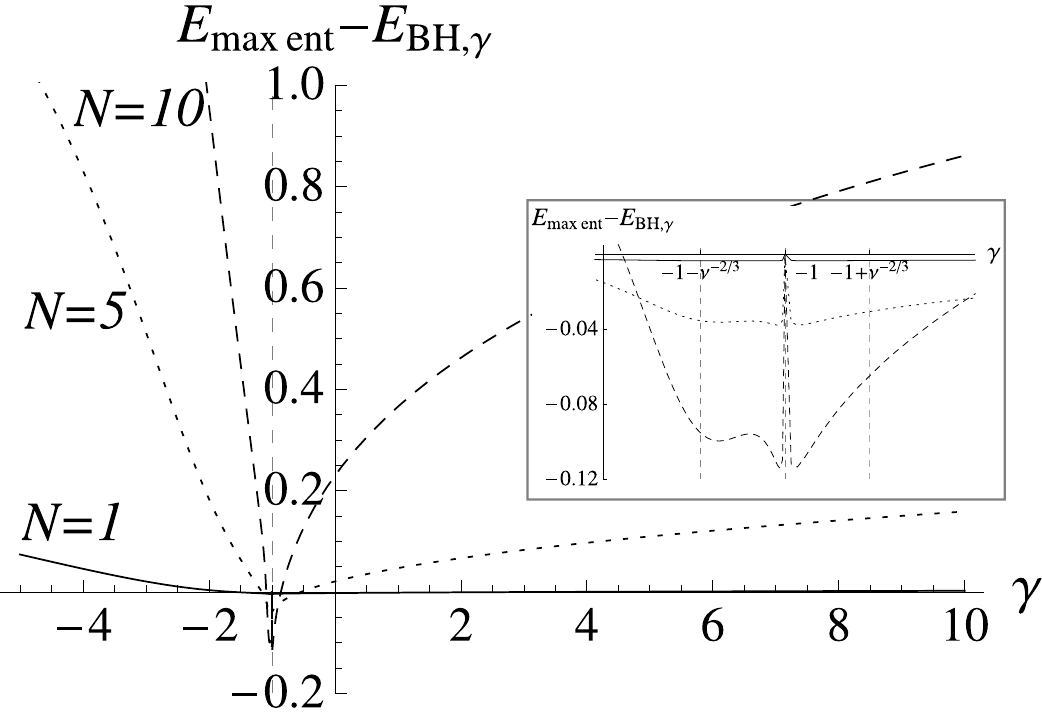}
}
\caption{Average final entanglement $E_{{\rm BH},-0.5}$, equation \eqref{av.ent}, of the teleportation protocol with the resource state being the ground state of the Hamiltonian (\ref{BH}) in the single Gaussian regime for $\nu=100$, equations \eqref{gs} and \eqref{gauss}. Panel (a) shows the average finale entanglement $E_{{\rm BH},-0.5}$, and panel (b) is the difference $E_{\rm max \, ent}-E_{{\rm BH},-0.5}$ with respect to the fidelity with the maximally entangled resource state \eqref{max.ent.state}, for $N=1$ (continuous line), $N=5$ (dotted line), and $N=10$ (dashed line). The insets show the zoom of the critical region $\left|\gamma+1\right|\leqslant\nu^{-2/3}$. Since the ground state of \eqref{BH} is not known in the critical region, the resource state of the teleportation fidelity plotted in the above insets is not the ground state of \eqref{BH}. It is instead the continuation of the ground state from the outside region $\left|\gamma+1\right|\geqslant\nu^{-2/3}$.}
\label{ent.BH}
\end{figure}
\end{widetext}

This concludes our analysis of the teleportation protocol described in section \ref{twomode}. In the next two sections we discuss generalizations of this protocol and the corresponding performances.

\section{Comments on the case of many modes} \label{multimode}

In this section we generalize the teleportation protocol by the use of many-mode states. As an introductory scenario, let us replace the first mode with a set of $m$ modes. The initial state is

\begin{equation} \label{initial.state.modes}
|\psi_{12}\rangle=\sum_{k=0}^N\sum_{\sigma=1}^{D_k^{(m)}}c_{k\sigma}|k,\sigma\rangle_1\otimes|N-k\rangle_2,
\end{equation}

\noindent
where $D_k^{(m)}={k+m-1 \choose k}$, and $\sigma$ is an additional index that distinguishes different orthogonal occupations of the first $m$ modes with $k$ particles. Since the goal is to teleport the single mode labelled by 2, we apply a two-mode resource state with $\nu$ particles (\ref{res}), and the teleportation protocol discussed in section \ref{twomode}, which does not act on the first $m$ modes. The average teleported state and the average fidelity are, respectively,

\begin{widetext}
\begin{eqnarray}
\label{teleported.modes} & & {\cal T}_m\big[|\psi_{12}\rangle\langle\psi_{12}\big]=\sum_{l=-N}^\nu\sum_{k,j=\max\{0,-l\}}^{\min\{N,\nu-l\}}\sum_{\sigma=1}^{D_k^{(m)}}\sum_{\tau=1}^{D_j^{(m)}}(\rho_{34})_{k+l,j+l} \, c_{k\sigma}\bar c_{j\tau}|k,\sigma\rangle_1{\,}_1\langle j,\tau|\otimes|N-k\rangle_4{\,}_4\langle N-j|, \\
\label{fidelity.modes} & & f_m=\frac{1}{D(D+1)}\sum_{l=-N}^\nu\left(\sum_{k,j=\max\{0,-l\}}^{\min\{N,\nu-l\}}D_k^{(m)}D_j^{(m)}(\rho_{34})_{k+l,j+l}+\sum_{k=\max\{0,-l\}}^{\min\{N,\nu-l\}}D_k^{(m)}(\rho_{34})_{k+l,k+l}\right).
\end{eqnarray}
\end{widetext}
The fidelity $f_m$ is straightforwardly computed insering the initial state \eqref{initial.state.modes} and the average teleported state \eqref{teleported.modes} into the general equation \eqref{fid}, and using averages of the probability distribution \eqref{measure} with $N+1$ replaced by $D$, being the Hilbert space dimension \eqref{dim.gen} in the present case.

\begin{widetext}

\begin{figure} [htbp]
\centering
\subfigure[]{
\label{fid.modes.a}
\includegraphics[width=0.48\columnwidth]{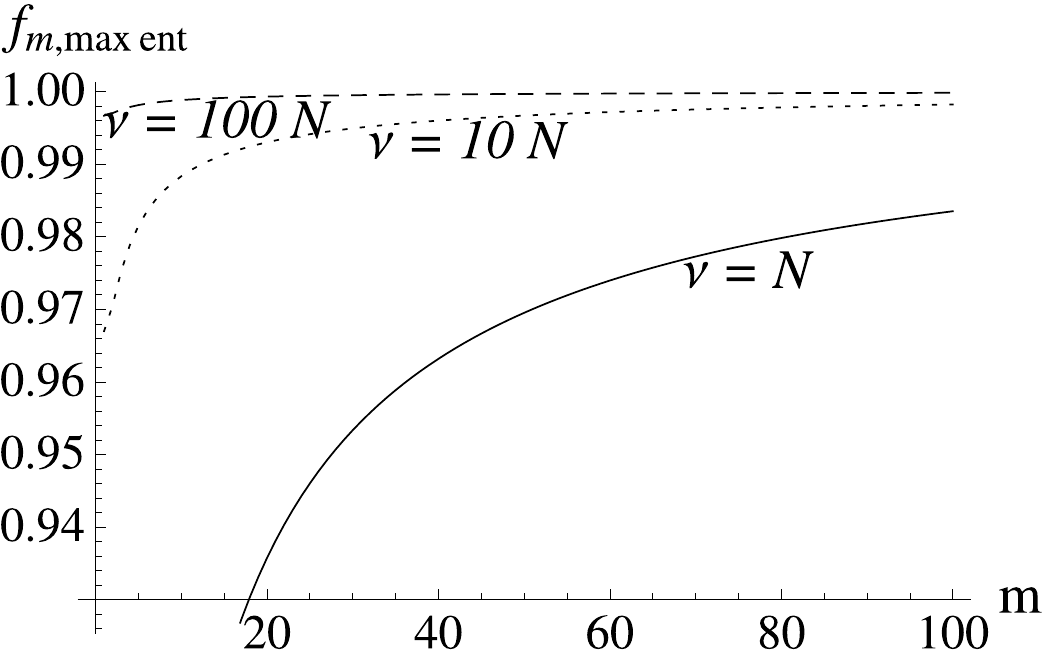}
}
\subfigure[]{
\label{fid.modes.b}
\includegraphics[width=0.48\columnwidth]{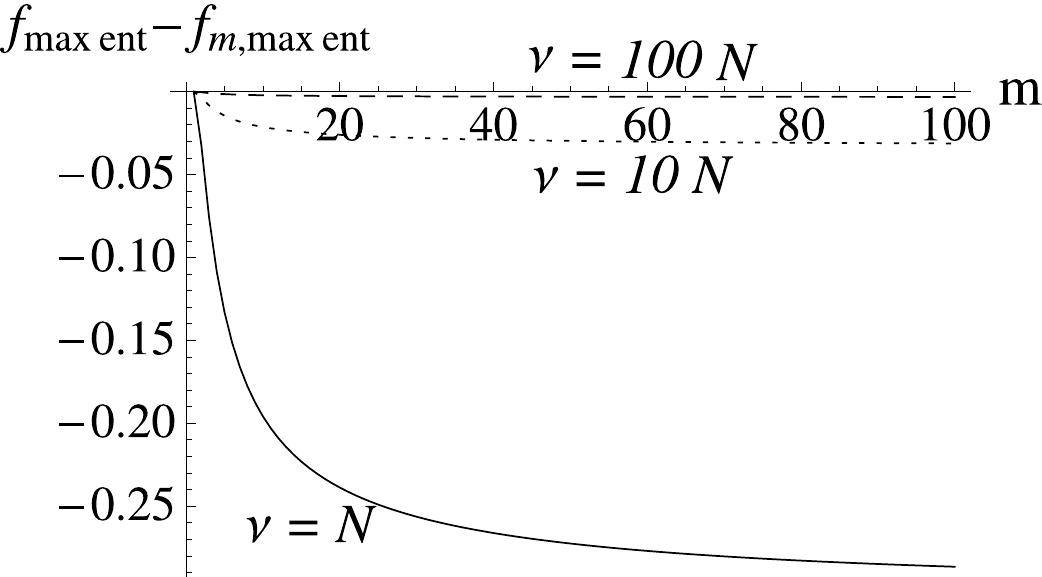}
}
\caption{Fidelity $f_{m,{\rm max \, ent}}$, equation \eqref{fidelity.modes}, of the teleportation protocol as function of the number of the first $m$ modes of the initial state \eqref{initial.state.modes}. Panel (a) shows the fidelity $f_{m,{\rm max \, ent}}$, and panel (b) the difference $f_{\rm max \, ent}-f_{m,{\rm max \, ent}}$ with $N=10$ particles in the initial state and $\nu=N$ (continuous line), $\nu=10N$ (dotted line), $\nu=100N$ (dashed line). The teleportation fidelity $f_{m,{\rm max \, ent}}$ monotonically increases with $m$.}
\label{fid.modes}
\end{figure}
\end{widetext}

In figure \ref{fid.modes}, we plot the fidelity $f_{m,{\rm max \, ent}}$ for the maximally entangled resource state (\ref{max.ent.state}), and its difference with respect to the fidelity $f_{\rm max \, ent}$ of the same protocol with $m=1$ discussed above. We consider the special case of $N=10$ particles in the state to be teleported, and notice that the fidelity increases with $m$. This is an effect of the dimensionality of the subsystem $1$. Thus, as soon as $m>1$, the fidelity is always larger than the fidelity with $m=1$. Furthermore, Alice measures the pairs $(l,\lambda)$ which allow a perfect teleportation with probability $\frac{\nu-N+1}{\nu+1}$. The fidelity for the maximally entangled resource states is bounded from below by such probability. Figure \ref{fid.modes} shows that this bound is not saturated in general, since the fidelity $f_{m,{\rm max \, ent}}$ can be larger than $f_{\rm max \, ent}$ that is in turn larger than the probability of a perfect teleportation. Moreover, the fidelity increases with the number $\nu$ of particles in the resource state. This can be explained with the inceasing amount of coherence and entanglement, which reduces more and more the probability of imperfect teleportation, as happens for two-mode initial states.

Let us now move to a more general setting. Alice wants to teleport a set $G_2$ of modes, which are entangled with another set $G_1$ of modes in the initial state $|\psi_{G_1,G_2}\rangle$. The shared resource state is $\rho_{G_3,G_4}$, where $G_3$ labels the set of modes owned by Alice, and $G_4$ labels the set of modes owned by Bob. Alice can teleport each mode of $G_2$ one at a time, applying the protocol described in section \ref{twomode}. High fidelity is achieved if the two agents share a maximally entangled two-mode state with $\nu$ particles for each mode in the set $G_2$. Therefore, if there are $m$ modes in each of the sets $G_2$, $G_3$ and $G_4$, the resource state is $\rho_{G_3,G_4}=|\phi_{G_3,G_4}\rangle\langle\phi_{G_3,G_4}|$, where

\begin{equation} \label{many.two.mode}
|\phi_{G_3,G_4}\rangle=\bigotimes_{j=1}^m\frac{1}{\sqrt{\nu+1}}\sum_{k=0}^\nu|k\rangle_{j^{(3)}}\otimes|\nu-k\rangle_{j^{(4)}},
\end{equation}

\noindent
and $j^{(3)}$ and $j^{(4)}$ label the modes in the set $G_3$ and $G_4$, respectively. The fidelity of this protocol is bounded from below by $\left(\frac{\nu-N+1}{\nu+1}\right)^m$, which is the probability that each mode is perfectly teleported. Figure \ref{fid.modes} shows that this is not a sharp bound, since the fidelity of the teleportation of each single mode can be larger than this lower bound, as discussed above. The lower bound, and thus the fidelity, are arbitrarily close to one if $\nu\gg N$ and $m$ is finite.

The state (\ref{many.two.mode}) is not a maximally entangled state between the sets of modes $G_3$ and $G_4$, because the reduced states are not completely mixed. It is however an extremely useful resource for the just mentioned procol. A natural question is whether we can achieve a higher fidelity by teleporting all the modes in $G_2$ in one single run. Such a protocol would be the straightforward generalization of the protocol with two-mode resource states: \emph{i)} Alice performs a projective measurement including projectors onto highly entangled states between $G_2$ and $G_3$, then \emph{ii)} she communicates the outcome to Bob, and finally \emph{iii)} Bob transforms his modes according to the received outcome.

Let us first consider the simpler case of an initial state $|\psi_{G_1,G_2}\rangle$ lying on one diagonal block (\ref{block.diag}), i.e. with a fixed local number $k$ of particles in the modes $G_1$ and $N-k$ in the modes $G_2$. The subspace of these states is isomorphic to the unconstrained tensor product space $\mathbbm{C}^{D_k^{(m)}}\otimes\mathbbm{C}^{D_{N-k}^{(M-m)}}$, and thus to the space of two distinguishable systems of dimensions $D_k^{(m)}$ and $D_{N-k}^{(M-m)}$, respectivelty. Hence, the set of modes $G_2$ can be perfectly teleported by the usual teleportation protocol \cite{Bennett1993} for a $D_{N-k}^{(M-m)}$-level system, translated to the formalism of Fock space. To this aim, we need a maximally entangled resource state between two $D_{N-k}^{(M-m)}$-level systems. This is realized by a pure state $|\phi_{G_3,G_4}\rangle$, where both $G_3$ and $G_4$ are sets of $M-m$ modes with a fixed number $N-k$ of particles in each of them. This protocol turns out to be exactly the same as discussed in \cite{Marinatto2001,Molotkov2010}, once the local number of particles and the number of modes are set, respectively, to $k=N-k=1$ and $m=M-m=2$. This latter example has been derived within the formalism of first quantization \cite{Marinatto2001,Molotkov2010}, and seems a bit intricate because of the permutation invariance of states. On the other hand, in second quantization we only need to straightforwardly apply the usual teleportation protocol, because the symmetrization is implicitly included in the formalism.

Now, we can wonder whether we can teleport all the modes in $G_2$ in one single run even for non-block-diagonal initial states. Generalizing the protocol in section \ref{twomode}, Alice performs a projective measurement on the modes in $G_2$ and $G_3$. The measurement projects onto states that generalise \eqref{states.meas} to the case of many-mode subsystems. Some of these states are $N$-particle maximally entangled with respect to the bipartition $(G_2,G_3)$, in analogy to the two-mode states (\ref{states.meas}) with $l=0$, and other states have more than $N$ particles and the same amount of entanglement than the previous ones, generalising the two-mode states (\ref{states.meas}) with $l\neq 0$. Two main difficulties arise in this protocol, which decrease the fidelity and forbid a perfect teleportation. The first comes from the mismatch between the dimension of the reduced states of $G_2$ and $G_3$ and the number of orthogonal projectors onto states with the same amount of entanglement, as happens for the two-mode states where we completed the measurement with additional projectors (\ref{states.meas.2}).

The second difficulty concerns the non-existence of a complete orthonormal basis of maximally entangled pure states with a fixed total number of identical particles. Such a basis always exists for distinguishable particles and unconstrained Hilbert spaces \cite{Vollbrecht2000}. It exists for the Fock space of $N$ two-mode particles, for the Fock space of one $M$-mode particle symmetrically halved in two mode-partitions ($m=M/2$), and in the special case of $N=2$ particles that fill $M=4$ modes divided into two equal parties ($m=2$) \cite{ZippilliComm}. On the other hand, a straightforward computation shows that a complete basis of maximally entangled states does not exist for one particle in $M$ modes which are split into two unequal sets ($m\neq M/2$), or for $N=2$ $(M=3)$-mode particles. It is not clear whether the existence of such bases is related to the symmetry of the algebraic bipartition. The impossibility of such a complete orthonormal basis reduces the number of projectors in Alice's measurement, that generalise (\ref{states.meas}) and allow for a perfect teleportation.

\section{Teleportation with reference frames} \label{reference}

We showed that the performance of the above teleportation protocol with two-mode states is limited by the conservation of the total number of particles. The limitation comes from the impossibility of a complete projective measurement such that each outcome provides perfect teleportation. This in turn stems from the fact that the second and the third mode are entangled with the other modes, thus their reduced state does not have a fixed number of particles, while Alice's projectors do. One can wonder whether there is a different protocol that provides perfect teleportation. Possible generalizations allow many-mode states, different perhaps non-projective measurements on Alice's side, and the use of a general resource state. A more interesting generalization is to relax the constraint of total particle number conservation. Nevertheless, in this section, we will prove that the teleportation fidelity is never one, for any general teleportation protocol using a finite number of particles.

Before discussing this statement, we recall the basic idea of relaxing conservation laws in quantum information protocols. The presence of a conservation law is formulated in terms of a superselection rule, namely the requirement that all the physically addressable states and observables commute with the conserved quantities. This requirement was connected to the lack of a reference frame, as reviewed in \cite{Bartlett2007}. For instance, any operator $X$ compatible with the conservation of the total particle number is invariant under the twirling operation $X=\int_0^{2\pi}\frac{d\phi}{2\pi}e^{i\phi\hat N}X e^{-i\phi\hat N}$, where $\hat N$ is the total number operator. This means that the phase $\phi$ is not observable if the particle number is conserved, and that the physical operators are uniformly averaged over the phase shift $e^{i\phi\hat N}$. On the contrary, if the different values of $\phi$ can be distinguished with respect to a phase reference, the physical operators are no longer invariant under the twirling operation.

The standard method to bypass the constraints imposed by the superselection rule \cite{Bartlett2007} involves embedding the original system into a larger one, allowing interactions with a reservoir. The conservation law applies only to the total system, while the original system is much less constrained. We then apply the invariance of physical operators under twirling operation with respect to the number of particles $\hat N_S$ of the systems plus that of the reservoir, $\hat N_R$: $X=\int_0^{2\pi}\frac{d\phi}{2\pi}e^{i\phi(\hat N_S+\hat N_R)}X e^{-i\phi(\hat N_S+\hat N_R)}$. Equivalently, the phase $\phi$ conjugated to the total number of particles of the system plus the reservoir is not observable. Nevertheless, since the relative particle number $\hat N_S-\hat N_R$ is not a conserved quantity, the local phase conjugated to $\hat N_S$ can be distinguished by measurements of the reservoir that, thus, provides a quantum phase reference. To give a specific example, the two-mode state $\sum_{k=1}^n\alpha_k|k\rangle_1|k\rangle_2$ and the operator $a_1^\dag a_2^\dag$ do not satisfy the particle number superselection rule, while the three-mode state $\sum_{k=1}^n\alpha_k|k\rangle_1|k\rangle_2|2n-2k\rangle_3$ and the operator $a_1^\dag a_2^\dag a_3^2$ do, and reproduce the same statistics of the previous two-mode case. A natural question is whether this approach allows to overcome all the restrictions imposed by the superselection rule, such as the impossibility of perfect teleportation shown in Proposition \ref{prop1} and in section \ref{perf}. For instance, any unconstrained operations on the original systems can be mimicked by suitable operations on the total system \cite{Kitaev2004}. This consideration was used to prove that any unconstrained quantum protocol consisting of local unitary operations can be mimicked, controlling the interactions with local reservoirs \cite{Kitaev2004}. This result is not directly applicable to general processes which consist of local operations and classical communication (LOCC). In the present paper, we are interested in the possibility of mimicking the unconstrained teleportation, which occurs with fidelity one, by means of interactions with a reservoir.

We explicitely show that the teleportation fidelity is strictly smaller than one, for any teleportation protocol consisting of finitely many particles and of LOCC on the system and a resource state possibly correlated with a reservoir.

\begin{proposition} \label{prop.imp}
Deterministic perfect teleportation, namely with fidelity one, is never possible for a fixed and finite number of identical particles.
\end{proposition}

For the rest of this section we prove the above proposition. We start with a special generalized teleportation protocol that captures the salient features of the impossibility of perfect teleportation, and will discuss all possible extensions later. Consider sets $G_j$ with a number $|G_j|$ of modes and $j=1,\dots,5$. Using the orthonormal basis \eqref{sep.basis}, the generalization of the initial state (\ref{initial.state}) is

\begin{equation} \label{initial.state.reserv}
|\psi_{G_1,G_2}\rangle=\sum_{k=0}^N\sum_{\sigma=1}^{D_k^{(|G_1|)}}\sum_{\tau=1}^{D_{N-k}^{(|G_2|)}}c_{k,\sigma,\tau}|k,\sigma\rangle_{G_1}\otimes|N-k,\tau\rangle_{G_2},
\end{equation}
where $G_1$ ($G_2$) is a set of modes that generalizes the first (second) mode of the protocol in section \ref{twomode}, and we aim to teleport the set of modes $G_2$. Consider pure resource states of $\nu$ four-mode particles

\begin{widetext}
\begin{equation} \label{res.reserv}
|R_{G_3,G_4,G_5}\rangle=\sum_{s,t=0}^\nu\sum_{\zeta=1}^{D_s^{(|G_3|)}}\sum_{\eta=1}^{D_{t}^{(|G_4|)}}\sum_{\theta=1}^{D_{\nu-s-t}^{(|G_5|)}}\beta_{s,t,\zeta,\eta,\theta}|s,\zeta\rangle_{G_3}\otimes|t,\eta\rangle_{G_4}\otimes|\nu-s-t,\theta\rangle_{G_5}.
\end{equation}
\end{widetext}

\noindent
$G_3$ is a set of modes owned by Alice, while $G_4$ and $G_5$ are the sets of modes owned by Bob. We consider resource states that can be coupled to a shared particle reservoir which is necessary for mimicking a general operation unconstrained by the particle number superselection rule. The part of the reservoir possessed by Alice is included in the set $G_3$. Considering arbitrarily large sets of modes, there is neither physical reason nor notational convenience to identify the reservoir from the modes in $G_3$. It is however convenient to divide the modes possessed by Bob into two sets $G_4$ and $G_5$, such that $G_4$ has the same number of modes as $G_2$. Thus, $G_4$ is the target of the teleportation, and $G_5$ plays the role only of the reservoir.

Alice performs a projective measurement in this extended setting, then communicates the result to Bob, and Bob performs a suitable operation on his modes.
In general, Alice projects onto an orthonormal basis of pure states $\{|\phi^{(\alpha)}_{G_2,G_3}\rangle\}_\alpha$ with $\mu\leqslant\nu$ particles, and $\alpha$ labels the elements of the orthonormal basis:

\begin{equation} \label{proj.reserv}
|\phi^{(\alpha)}_{G_2,G_3}\rangle=\sum_{j=0}^\mu\sum_{\pi=1}^{D_j^{(|G_2|)}}\sum_{\omega=1}^{D_{\mu-j}^{(|G_3|)}}\Phi_{j,\pi,\omega}^{(\alpha)}|j,\pi\rangle_{G_2}\otimes|\mu-j,\omega\rangle_{G_3}.
\end{equation}

\noindent
After the projection the state becomes

\begin{widetext}
\begin{eqnarray} \label{measured}
|\phi^{(\alpha)}_{G_2,G_3}\rangle\langle\phi^{(\alpha)}_{G_2,G_3}|\psi_{G_1,G_2}\rangle\otimes|R_{G_3,G_4,G_5}\rangle & = & |\phi_{G_2,G_3}^{(\alpha)}\rangle\otimes\sum_{k,t,\sigma,\tau,\zeta,\eta,\theta}c_{k,\sigma,\tau}\overline{\Phi_{N-k,\tau,\zeta}^{(\alpha)}} \, \beta_{\mu-N+k,t,\zeta,\eta,\theta}|k,\sigma\rangle_1\otimes|t,\eta\rangle_{G_4}\otimes \nonumber \\
&& \otimes|\nu-\mu+N-k-t,\theta\rangle_{G_5},
\end{eqnarray}
\end{widetext}

Afterwards, Alice communicates the result $\alpha$ of her measurement to Bob who performs an operation $V_{G_4,G_5}^{(\alpha)}$ on his modes. The operation is aimed to recover a state which maximizes the fidelity, such that the final state of the modes in $G_1$ and $G_4$ is as similar as possible to the initial state \eqref{initial.state.reserv}. If a protocol with fidelity one is possibile, namely perfect teleportation from the modes in $G_2$ to the modes in $G_4$, then the final state has to be pure and Bob's modes have to be factorized with the remaining modes, because these are features of the initial state \eqref{initial.state.reserv}. Therefore, $V_{G_4,G_5}^{(\alpha)}$ has to preserve the norm of the state (\ref{measured}). Comparing the state after the measurement \eqref{measured} and the initial state \eqref{initial.state.reserv}, in order to maximize the fidelity, the state of the set $G_4$ that multiplies $|k,\sigma\rangle_{G_1}$ should be transformed into $|N-k,\tau\rangle_{G_4}$. Since these properties must be satisfied for all the initial states, i.e. for all the coefficients $c_{k,\sigma,\tau}$, the operator $V_{G_4,G_5}^{(\alpha)}$ must transform the states

\begin{equation}
\sum_{t,\eta,\theta}\beta_{\mu-N+k,t,\zeta,\eta,\theta}|t,\eta\rangle_{G_4}\otimes|\nu-\mu+N-k-t,\theta\rangle_{G_5}
\end{equation}

\noindent
into

\begin{equation}
e^{i\varphi(\alpha,k,\sigma,\tau,\zeta)}\sqrt{\sum_{t,\eta,\theta}|\beta_{\mu-N+k,t,\zeta,\eta,\theta}|^2}|N-k,\tau\rangle_4\otimes|\chi(\nu-\mu)\rangle_6
\end{equation}

\noindent
for all $k$ and $\tau$, where $\varphi(\alpha,k,\sigma,\tau,\zeta)$ is an arbitrary phase, and $|\chi(\nu-\mu)\rangle_{G_5}$ is a state with $\nu-\mu$ particles in the modes $G_5$ which is independent on the other indices. This transformation should rely on correlations among the indices $\tau$, $\zeta$ and $\eta$ induced by the coefficients $\overline{\Phi_{N-k,\tau,\zeta}^{(\alpha)}}$ and $\beta_{\mu-N+k,t,\zeta,\eta,\theta}$. The result is

\begin{widetext}
\begin{eqnarray}
& & \Big(|\phi^{(\alpha)}_{G_2,G_3}\rangle\langle\phi^{(\alpha)}_{G_2,G_3}|\otimes V_{G_4,G_5}^{(\alpha)}\Big)|\psi_{G_1,G_2}\rangle\otimes|R_{G_3,G_4,G_5}\rangle=|\phi_{G_2,G_3}^{(\alpha)}\rangle\otimes\sum_{k,\sigma,\tau,\zeta}c_{k,\sigma,\tau} \, e^{i\varphi(\alpha,k,\sigma,\tau,\zeta)} \, \overline{\Phi_{N-k,\tau,\zeta}^{(\alpha)}}\cdot \nonumber \\
& & \cdot\sqrt{\sum_{t,\eta,\theta}\left|\beta_{\mu-N+k,t,\zeta,\eta,\theta}\right|^2} \, |k,\sigma\rangle_{G_1}\otimes|N-k,\tau\rangle_{G_4}\otimes|\chi(\nu-\mu)\rangle_{G_5}.
\end{eqnarray}

\noindent
After tracing out all modes but those in $G_1$ and $G_4$, we get the unnormalized state

\begin{equation} \label{final.reserv}
|\psi_{G_1,G_4}^{(\alpha)}\rangle=\sum_{k=0}^N\sum_{\sigma,\tau,\zeta}c_{k,\sigma,\tau} \, e^{i\varphi(\alpha,k,\sigma,\tau,\zeta)} \, \overline{\Phi_{N-k,\tau,\zeta}^{(\alpha)}}\sqrt{\sum_{t,\eta,\theta}\left|\beta_{\mu-N+k,t,\zeta,\eta,\theta}\right|^2} \, |k,\sigma\rangle_{G_1}\otimes|N-k,\tau\rangle_{G_4}.
\end{equation}
\end{widetext}

From its definition \eqref{fidelity}, the average fidelity of the teleportation can be recast into

\begin{equation} \label{fid.reserv}
f=\int d\psi\sum_{\alpha}\langle\psi_{G_1,G_4}^{(\alpha)}|\psi_{G_1,G_4}^{(\alpha)}\rangle\frac{\left|\langle\psi_{G_1,G_4}|\psi_{G_1,G_4}^{(\alpha)}\rangle\right|^2}{\langle\psi_{G_1,G_4}^{(\alpha)}|\psi_{G_1,G_4}^{(\alpha)}\rangle},
\end{equation}

\noindent
where $|\psi_{G_1,G_4}\rangle$ is the same as the initial state \eqref{initial.state.reserv} with the only difference that the set of modes $G_2$ is replaced by $G_4$. Equation \eqref{fid.reserv} is the average of $\frac{\left|\langle\psi|\psi^{(\alpha)}\rangle\right|^2}{\langle\psi^{(\alpha)}|\psi^{(\alpha)}\rangle}$ with probability $\langle\psi^{(\alpha)}|\psi^{(\alpha)}\rangle$. Assuming that the fidelity is one, the overlap between any initial state and the normalized state resulting from the $\alpha$-th outcome should be one, namely

\begin{equation} \label{fid.part}
\frac{\left|\langle\psi|\psi^{(\alpha)}\rangle\right|^2}{\langle\psi^{(\alpha)}|\psi^{(\alpha)}\rangle}=\frac{\displaystyle \left|\sum_{k,\sigma,\tau}|c_{k,\sigma,\tau}|^2 \gamma_{k,\tau}^{(\alpha)}\right|^2}{\displaystyle \sum_{k,\sigma,\tau}|c_{k,\sigma,\tau}|^2|\gamma_{k,\tau}^{(\alpha)}|^2}=1,
\end{equation}
with

\begin{equation}
\gamma_{k,\tau}^{(\alpha)}=\sum_\zeta e^{i\varphi(\alpha,k,\sigma,\tau,\zeta)} \, \overline{\Phi_{N-k,\tau,\zeta}^{(\alpha)}}\sqrt{\sum_{t,\eta,\theta}\left|\beta_{\mu-N+k,t,\zeta,\eta,\theta}\right|^2},
\end{equation}

\noindent
for all the initial states $|\psi\rangle$. Recalling the notation \eqref{sep.basis}, and considering, among all possible initial states \eqref{initial.state.reserv},

\begin{equation} \label{initial.particular}
|\psi\rangle=c \, |k,\sigma\rangle|N-k,\tau\rangle+\sqrt{1-c^2} \, |k',\sigma'\rangle|N-k',\tau'\rangle,
\end{equation}
with $0<c<1$, the condition (\ref{fid.part}) can be re-written, after some simple algebra, as

\begin{equation}
(c^2-c^4)|\gamma_{k,\tau}^{(\alpha)}-\gamma_{k',\tau'}^{(\alpha)}|^2=0.
\end{equation}
The latter equation holds for all $k$, $k'$, $\tau$ and $\tau'$ that can be arbitrary chosen in the exemplary initial state \eqref{initial.particular}, and thus implies $\gamma_{k,\tau}^{(\alpha)}=\gamma^{(\alpha)}$ are indendent on $k$ and $\tau$. Plugging this result into the definition of the fidelity, we get $f=\sum_\alpha|\gamma^{(\alpha)}|^2$. The fidelity is maximized if the cardinality of the sum over $\alpha$ is maximal, that corresponds to a projective measurement onto the complete orthonormal basis $\{|\phi_{G_2,G_3}^{(\alpha)}\rangle\}_\alpha$. The completeness of the measurement implies the identity

\begin{equation}
\sum_\alpha \overline{\Phi_{j,\pi,\omega}^{(\alpha)}}\Phi_{j',\pi',\omega'}^{(\alpha)}=\delta_{j,j'}\delta_{\pi,\pi'}\delta_{\omega,\omega'}.
\end{equation}
Hence, a perfect teleportation implies

\begin{eqnarray} \label{contradiction}
1 & = & \sum_{\alpha}|\gamma^{(\alpha)}|^2=\sum_{t,\zeta,\eta,\theta}|\beta_{\mu-N+k,t,\zeta,\eta,\theta}|^{2} \nonumber \\
&& <\sum_{s,t,\zeta,\eta,\theta}|\beta_{s,t,\zeta,\eta,\theta}|^{2}=1
\end{eqnarray}

\noindent
which is a contradiction.

The only assumption we made is that the teleportation is perfect, namely that the fidelity is one. Therefore, a perfect teleportation cannot be implemented exploiting particle reservoirs. If each set $G_{1,2,3,4}$ is made of one mode and the modes in $G_5$ are factorized from the others in the resource state (\ref{res.reserv}) and in the projectors (\ref{proj.reserv}), the proof recovers a generalization of the teleportation in section \ref{twomode} without reservoirs. In the following, the proof is generalized along several directions in order to recover all possible teleportation protocols described by LOCC.

\begin{itemize}
\item We only considered pure resource states because the maximum fidelity is attained by a pure resource state. Indeed, the linearity of teleportation implies that the fidelity provided by a mixture of pure resource states is the convex combination of the fidelities provided by each pure resource state. Thus, if the fidelity is strictly smaller than one for pure resource states, it is as well for mixed resource states.

\item The above proof can be generalized to a scenario where Alice projects onto pure states with different total numbers of particles and when the resource state is a mixture of states with different total numbers of particles. Indeed, the operation $V_{G_4,G_5}^{(\alpha)}$ can be optimized only for one choice of the couple $(\mu,\nu)$, introducing addional errors for the other values.

\item We can further generalize the argument when Alice projects onto degenerate subspaces. In this case, the modes in $G_2$ and $G_3$ will be entangled with the other modes, providing a decrease of the fidelity when they are traced out.

\item There is no loss of generality in considering projective measurements on Alice's side. In fact, it is known \cite{NielsenChuang} that any quantum operation, e.g a generalized measurement or completely positive map, is equivalent to a unitary operation $U$ followed by a projective measurement on an enlarged system. Thus, unitary operations $U$ on Alice's side do not change Bob's final state, and the extension of the system can be gathered in the set $G_3$. Hence, this more general scheme is absorbed by the previous case.

\item If no projective measurements are involved, there is no broadcast of information to Bob's end. When a projective measurement is performed, information is teleported to Bob's modes. The cost of the projection is the complete erasure of information on Alice's side. For this reason, no further iterations or backward communication from Bob to Alice can improve the teleportation fidelity. This proves the statement for a general LOCC protcol.
\end{itemize}

There is still the possibility to consider an infinite number of particles. We have already shown that such an asymptotic, perfect teleportation is possible with the teleportation protocol described in section \ref{twomode}, without any further generalization.

\section{Conclusions} \label{discussions}

We have discussed how the teleportation protocol can be applied to the case of identical massive particles. Due to the indistinguishability of particles, we applied a general notion of entanglement between subalgebras of observables. As a consequence, we identified local parties with orthogonal modes, such as in optical lattices where we can split the wells into groups. We considered the following situation: a sender, Alice, wants to teleport the state of one of her modes to a mode owned by a receiver, Bob. To this aim, they use an entangled shared two-mode state. In general, the mode whose state Alice wants to teleport can be entangled with another mode.

We computed the general formula of the teleportation fidelity, that is the average overlap between the initial pure state and the teleported state, and the average entanglement of the teleported state, when one mode of a two-mode state is teleported. We proved that the conservation of the total number of particles forbids the fidelity to be one for a finite number of particles, even if coherent interactions with a reservoir are allowed to overcome the superselection rule. We computed the teleportation performances for several states. In particular, the maximally entangled state (\ref{max.ent.state}), the symmetric coherent state and the ground state of the two-mode Bose-Hubbard Hamiltonian (\ref{BH}) provide perfect teleportation in the limit of large numbers of particles. Each of these states is interesting for different reasons. The maximally entangled state (\ref{max.ent.state}) provides perfect teleportation with high probability for a finite number of particles. The symmetric coherent state can be prepared with current technologies in systems of ultracold atoms \cite{Gross2010,Riedel2010,Diehl2008}. Moreover, it is closly related to the mean field approximation and mesoscopic quantum coherent phenomena \cite{Raghavan1999,Benatti2009}, and is considered a classical state for metrological purposes \cite{Wineland1994,Benatti2010,Benatti2011,Argentieri2011,Benatti2014}, while it is very useful for teleportation. In comparison, N00N states are extremely useful in quantum metrology \cite{Bollinger1996,Hyllus2010} while they are not for teleportation. These differences are a consequence of the fact that teleportation and metrological performaces require different state properties: large entanglement, thus coherence among all the Fock state, for teleportation, and coherence between two Fock states with highly unbalanced population in two modes for phase estimation. The teleportation performances of the ground state of the double well potential are appealing because this state can be generated with available techniques, i.e. magnetic traps and evaporative cooling \cite{Thomas2002}. Finally, we briefly discussed possible generalizations to the teleportation of many modes and related difficulties in achieving high performances.

{\bf Acknowledgement} U.M. acknowledges Stefano Zippilli for useful discussions, and funding by the grant J1-5439 of Slovenian Research Agency. U.M. and A.B. acknoledge funding by Deutsche Forschungsgemeinschaft and by Evaluierter Fonds der Albert Ludwigs-Universitaet Freiburg.

\appendix

\section{Teleportation performances with exemplary resource states} \label{perf.app}

In this appendix, we describe some additional details on teleportation performaces with the resource states discussed in section \ref{perf}.

\subsection{Separable resources}

When a separable resource state \eqref{sep.res} is considered, the average teleported state is

\begin{widetext}
\begin{eqnarray} \label{teleported.sep}
\mathcal{T}_{\rm sep}\big[|\psi_{12}\rangle\langle\psi_{12}|\big]=\sum_{l=-N}^\nu\sum_{k=\max\{0,-l\}}^{\min\{N,\nu-l\}} |c_k|^2\left(\rho_{34}\right)_{k+l,k+l}|k\rangle_1{\,}_1\langle k|\otimes|N-k\rangle_4{\,}_4\langle N-k|,
\end{eqnarray}
\end{widetext}

\noindent
which is diagonal in the Fock basis, because, after Alice's measurement, the state \eqref{telep.term} is always separable. The teleportation fidelity and the average teleported entanglement are given respectively by equation \eqref{fid.sep} and $E_{\rm sep}=0$.

The inequality \eqref{triangle} is equivalent to

\begin{equation}
E\geq\frac{\pi}{8}(N+2)(f-f_{\rm sep}).
\end{equation}
Thus, if the fidelity of a resource state overcomes the fidelity of separable resource states $f>f_{\rm sep}$, the same relation holds for the average final entanglement $E>E_{\rm sep}=0$.

\subsection{Maximally entangled resources and probabilistic, perfect teleportation}

In order to analyse the performance of the maximally entangled resource state \eqref{max.ent.state}, define the projectors

\begin{equation} \label{proj.Q}
Q_l=\sum_{k=\max\{0,-l\}}^{\min\{N,\nu-l\}}|k\rangle_1{\,}_1\langle k|\otimes|N-k\rangle_4{\,}_4\langle N-k|,
\end{equation}

\noindent
and $|\psi_{14}\rangle$ the perfectly teleported state, i.e. the same state as \eqref{initial.state} but pertaining to the first and fourth mode. The average teleported state is

\begin{widetext}
\begin{equation}
\mathcal{T}_{\rm max \, ent}\big[|\psi_{12}\rangle\langle\psi_{12}|\big]=\frac{\nu-N+1}{\nu+1}|\psi_{14}\rangle\langle\psi_{14}|+\frac{1}{\nu+1}\left(\sum_{l=-N}^{-1}Q_l|\psi_{14}\rangle\langle\psi_{14}|Q_l+\sum_{l=\nu-N+1}^{\nu}Q_l|\psi_{14}\rangle\langle\psi_{14}|Q_l\right).
\end{equation}
\end{widetext}

\noindent
If Alice's measurement results in a projection onto \eqref{states.meas.2} with a specific $(l,\lambda)$, the component $Q_l|\psi\rangle$ of the initial state \eqref{initial.state} is teleported. This projection exactly recovers the initial state, if $0\leqslant l\leqslant\nu-N$, resulting in a perfect teleportation, namely the mapping from the second mode to the fourth mode, which occurs with probability $\frac{\nu-N+1}{\nu+1}$. All the other outcomes result in the partial teleportation of the component $Q_l|\psi_{14}\rangle$, with probability $\frac{1}{\nu+1}\langle\psi_{14}|Q_l|\psi_{14}\rangle$. This lead to teleportation fidelity \eqref{fid.max.ent} and to average maximally entanglement \eqref{ent.max.ent}.

A possible strategy to optimize the performance of the teleportation protocol is to increase the probability of the perfect teleportation. On the one hand, such probability $p_{{\rm perf}}^{(1)}=\frac{\nu-N+1}{\nu+1}$ increases with the ratio $\nu/N$ and goes to one when $\nu/N\to\infty$. This requires the ability to prepare states \eqref{max.ent.state} for a very large number of particles $\nu$. On the other hand, considering a fixed and finite number of particles $\nu$, we can improve the probability of perfect teleportation with repeated teleportations. If Alice's projection onto \eqref{states.meas.2} corresponds to a value $l<0$ or $l>\nu-N$, which does not allow a perfect teleportation, she teleports another copy of the original state until she gets the outcome $0\leqslant l\leqslant\nu-N$, namely a perfect teleportation. After $r$ runs of the teleportation protocol, the probability of perfect teleportation is

\begin{eqnarray}
p_{{\rm perf}}^{(r)} & = & \sum_{m=0}^{r-1}(1-p_{{\rm perf}}^{(1)})^m p_{{\rm perf}}^{(1)}=1-(1-p_{{\rm perf}}^{(1)})^r \nonumber \\
& = & 1-\left(\frac{N}{\nu+1}\right)^r.
\end{eqnarray}

Any other maximally entangled state \cite{Benatti2012,Benatti2012-2}

\begin{equation}
|\tilde\phi_{34}\rangle=\frac{1}{\sqrt{\nu+1}}\sum_{k=0}^\nu e^{i\vartheta(k)}|k\rangle_3\otimes|\nu-k\rangle_4,
\end{equation}

\noindent
with arbitrary phases $\vartheta(k)$ can be transformed to $|\phi_{34}\rangle$ by means of a local unitary operation, e.g.

\begin{equation} \label{abs.unit}
\sum_{k=0}^N e^{-i\vartheta(k)}|k\rangle_4{\,}_4\langle k|\tilde\phi_{34}\rangle=e^{-i\vartheta(a_4^\dag a_4)}|\tilde\phi_{34}\rangle=|\phi_{34}\rangle.
\end{equation}

\noindent
Therefore all the maximally entangled resource states provide the same performance, up to local unitary operations. These local unitary operations can be reabsorbed in the protocol, for instance redefining $V_4^{(l,\lambda)}$. The choice in \eqref{V_4} maximizes the fidelity of the maximally entangled state \eqref{max.ent.state}, whereas it is not optimal for other maximally entangled states which can lead to a fidelity smaller than that of separable states. The redefinition of $V_4^{(l,\lambda)}$, by absorption of the unitaries \eqref{abs.unit}, does not affect the average final entanglement $E_{\rm max \, ent}$.

\subsection{N00N states}

The N00N states \eqref{N00N} discussed in section \ref{N00N.states} can be transformed into N00N states with $n\leqslant N$ particles, through local and controlled particle losses:

\begin{equation}
|n00n\rangle_{34}=\frac{1}{\sqrt{2}}\big(|n\rangle_3\otimes|0\rangle_4+|0\rangle_3\otimes|n\rangle_4\big)=W_3\otimes W_4|\nu00\nu\rangle_{34},
\end{equation}

\noindent
where

\begin{equation}
W_j=|0\rangle_j{\,}_j\langle 0|+|n\rangle_j{\,}_j\langle\nu|.
\end{equation}

\noindent
The N00N states with $n\leqslant N$ particles exhibit better teleportation performances than separable resource states:

\begin{eqnarray}
f_{n00n} & = & \frac{2}{N+2}\left(1+\frac{N-n+1}{2(N+1)}\right)<\frac{3}{N+2}, \label{fid.N00N} \\
E_{n00n} & = & \frac{\pi(N-n+1)}{8(N+1)}<\frac{\pi}{8}, \label{av.ent.N00N}
\end{eqnarray}
that are directly computed from the general equations \eqref{fidelity} and \eqref{av.ent}. The second term of \eqref{fid.N00N} is the additional contribution to the fidelity with respect to separable resources. It is negligible for large $n\simeq N$, increases with decreasing $n$, and contributes to the teleportation fidelity at the same order as $f_{\rm sep}$ \eqref{fid.sep}. Thus, the resulting improvement gives at most a larger prefactor, without changing the scaling with the numbers of particles. Moreover, the average final entanglement is of order one, while the maximum value of the negativity and its average over all pure states scale linearly with $N$.

We stress that the operations $W_j$ are fully consistent with the conservation of the total particle number, which plays a crucial role in our study. Indeed, $W_3$ can be implemented with a unitary operation on the third mode and on an additional fifth mode, to preserve the total particle number: $W_3\rho_3 W_3^\dag={\rm tr}_5\big(\tilde W_{35}\rho_3\otimes|0\rangle_5{\,}_5\langle 0|\tilde W_{35}^\dag\big)$, where ${\rm tr}_5$ is the trace over the fifth mode, and

\begin{eqnarray}
\tilde W_{35} & = & |0\rangle_3{\,}_3\langle 0|\otimes|0\rangle_5{\,}_5\langle 0|+\big(|n\rangle_3{\,}_3\langle \nu|\otimes|\nu-n\rangle_5{\,}_5\langle 0| \nonumber \\
&& +{\rm h.c.}\big)+\mathbbm{1}_{n,\nu}.
\end{eqnarray}

\noindent
$\mathbbm{1}_{n,\nu}$ is the identity matrix on the subspace orthogonal to the support of each of the other terms. Thus, $\tilde W_{35}$ is a unitary transformation which commutes with the total number of particles. Analogously, $W_4$ can be implemented with a total number preserving unitary operation on the fourth mode and an additional sixth mode.

\subsection{SU(2) coherent states}

In addition to the discussion in section \ref{SU2.coh.states} and from figures \ref{fid.sym.coh} and \ref{ent.sym.coh}, we note that the SU(2) symmetric coherent state $|\xi=1/2,\vartheta=0\rangle$ outperforms the maximally entangled states \eqref{max.ent.state} for $1\leqslant N\leqslant 3$ up to large particle numbers $\nu$. The intuitive reason is the following: If a maximally entangled state is employed as a resource, the teleportation is perfect whenever Alice measures $(l,\lambda)$ with $0\leqslant l\leqslant \nu-N$, which happens with a probability $\frac{\nu-N+1}{\nu+1}$. The teleported state resulting form any other of Alice's outcomes is a projection (\ref{proj.Q}) of the original state. For a symmetric coherent resource state, the teleportation corresponding to Alice's outcomes $(l,\lambda)$, with $0\leqslant l\leqslant \nu-N$, is slightly distorted with respect to the perfect teleportation, but these outcomes do occur with higher probability than in the case of the maximally entangled resource state.

As an example of entangled resource state that does not outperform separable states, we numerically checked that the teleportation fidelity of the coherent state $|\xi=1/2,\vartheta=\pi\rangle_{34}$ is smaller than the fidelity of separable states $f_{\rm sep}$, for $N\in[1,10]$ and $\nu\in[1,100]$. Indeed, from figure \ref{perf.coh} the maximum fidelity with SU(2) coherent resource states is achieved for the symmetric state $|\xi=1/2,\vartheta=0\rangle$.

Any coherent state is equivalent to any other coherent state with the same value of $\xi$ and different phases $\vartheta$, up to local unitary operations on the modes, i.e. of the form $e^{i\Theta(a_4^\dag a_4)}$ with a given function $\Theta(\cdot)$. This means that coherent states with the same $\xi$ but different $\vartheta$ have the same entanglement \cite{Vidal2000}. Nevertheless, these local operations can be reabsorbed in the operations $V_4^{(l,\lambda)}$, changing the teleportation protocol and thus its performance. Consistently, we find a dependence of the fidelity on the phase $\vartheta$, even if neither the entanglement of SU(2) coherent resource states nor the average final entanglement they produce change with $\vartheta$.

\subsection{Ground states of the double well potential}

In this section, we stress some features of teleportation performances with the resource state being the ground state of the Bose-Hubbard model \eqref{BH}. We numerically observed that the behaviour of the fidelity \eqref{fidelity} and of the average final entanglement \eqref{av.ent}, when the ground state is chosen in the double Gaussian regime, are qualitatively the same as in the single Gaussian regime. The reason is that for small (large) $\nu$ the Gaussians are highly peaked (strongly spread out), and the small (large) coherence among Fock states, thus the entanglement, justify low (high) teleportation performances. Increasing $\nu$, the spread of the Gaussians increases monotonically, as well as the teleportation performances. Unfortunately, there is no explicit formula for the ground state in the critical regime.

The fidelity $f_{{\rm BH},\gamma}$ and the average final entanglement $E_{{\rm BH},\gamma}$ are larger than the fidelity $f_{\rm max \, ent}$ and the average final entanglement $E_{\rm max \, ent}$ of the maximally entangled state, respectively, for small values of $N$ up to large $\nu$. The intuitive reason is the same as the one given above for the symmetric coherent state. The number of such values of $N$ increases when $\gamma$ approaches $-1$, and decreases to zero when $\gamma$ moves away from $-1$. In the absence of interactions, i.e. $\gamma=0$, the exact ground state of the Hamiltonian (\ref{BH}) is the symmetric coherent state \eqref{def.coh} with $\xi=1/2$ and $\vartheta=0$. Indeed, the symmetric coherent state is approximated by the Gaussian state (\ref{gauss}) with $\gamma=0$, as can be seen by application of Stirling's approximation \cite{Chandrasekhar1943} for a large number of particles $\nu$. Consistently, the fidelities and the average final entanglement of these resource states have very similar quantitative behaviours, like in Figures \ref{fid.sym.coh} and \ref{ent.sym.coh}.

\end{document}